\documentclass[aps,prb,twocolumn,showpacs,superscriptaddress]{revtex4}
\usepackage{epsfig}
\begin{document}
\title{
Thermoelectric properties of junctions
between metal and models of strongly correlated semiconductors}

\author{Massimo Rontani}
\email{rontani@unimore.it}
\homepage{http://www.nanoscience.unimo.it/max_index.html}

\affiliation{Department of Physics, University of California San Diego,
La Jolla, California 92093-0319}

\affiliation{INFM National Research Center on nanoStructures
and bioSystems at Surfaces (S3) and \\
Dipartimento di Fisica, Universit\`a degli Studi di Modena
e Reggio Emilia, 41100 Modena, Italy}

\author{L.~J.~Sham}

\affiliation{Department of Physics, University of California San Diego,
La Jolla, California 92093-0319}

\date{\today}
\begin{abstract}
We study the thermopower of a junction between a metal and a
strongly correlated semiconductor. Both in the electronic ferroelectric
regime and in the Kondo insulator regime the thermoelectric 
figures of merit, $ZT$, of these junctions are compared with
that of the ordinary semiconductor. By inserting at the interface
one or two monolayers of atoms different from the bulk, 
with a suitable choice of rare-earth elements very high values
of $ZT$ can be reached at low temperatures. The potential of
the junction as a thermoelectric device is discussed.
\end{abstract}
\pacs{72.10.Fk, 73.40.Cg, 73.40.Ns, 73.50.Lw}

\maketitle

\section{Introduction}

New thermoelectric coolers and power generators are under
massive investigation.\cite{physicstoday} 
These devices are in general even more reliable
than commercial heat-exchange refrigerators but their efficiency, 
in the best cases, is much lower. 
The quality of the material need in a thermoelectric device is 
defined by the dimensionless figure of merit
$ZT$, where $T$ is the absolute temperature, and $Z$ is
expressed in Eq.~(\ref{eq:defZT}) in terms of transport 
coefficients.
Currently, the highest value of $ZT$, which is $\sim$ 1 at room
temperature, is found in Bi-Te alloys.\cite{superrama} New devices would be 
competitive with traditional refrigerators if $ZT$ were about 3 
$\sim$ 4: the major lack of high-$ZT$ materials is at temperatures
below \protect{300 K}.

In this paper\cite{us} we propose a junction of metal and strongly 
correlated semiconductor as the basis for a possible efficient 
low-temperature thermoelectric device.  This system embodies previous 
intuitions that were recognized fruitful\cite{sales,procnat,optimumgap,mao,2d,2d2,2d3,mahan2d,nahum,edwards,thermionic,moyzhes,min} 
in different materials/devices 
such as rare-earth compounds, superlattices, and metal/superconductor 
junctions. We now outline these concepts and briefly review the related 
literature.

From the definition
\begin{equation}
Z=\frac{Q^2\sigma}{\left(\kappa_e+\kappa_l\right)},
\label{eq:defZT}
\end{equation}
where $Q$ is the absolute thermopower (Seebeck coefficient), $\sigma$
the electrical conductivity, and $\kappa_e$ and $\kappa_l$ the electronic and
lattice part, respectively, of the thermal conductivity, it
follows that an ideal thermoelectric material should have
high thermopower, high electrical conductivity, and low 
thermal conductivity. Semiconductors seemed to have
the optimum collection of these properties, in contrast with
metals, which have high $\sigma$ but low $Q$, and insulators, 
which have high $Q$ but low $\sigma$.
However, in the last thirty years no substantial enhancement of $ZT$ 
beyond $\sim$ 1 was obtained.

A breakthrough came with the synthesis of new materials such as 
filled skutterudites, 
which have nearly bound rare-earth atoms, closed in an atomic cage,
whose ``rattling'' under thermal excitation scatters phonons, 
then dramatically reducing $\kappa_l$.\cite{sales} 
More generally, heavy atoms in compounds help with lowering $\kappa_l$.
Mahan and Sofo\cite{procnat} also showed that the best bulk band structure 
for high $Q$ is one with a sharp singularity in the density
of states very close to the Fermi energy. These results provide
the first idea in the search for the best thermoelectric, 
namely to look at rare-earth compounds as major candidates.
In fact, mixed valence metallic compounds (e.g.~CePd$_3$, YbAl$_3$) show
high values of $Q$, but at the present time no useful
value of $ZT$ has been reported.\cite{rareearth}

The second idea is that the best thermoelectric must have an energy gap.
Because in ordinary semiconductors the optimum band gap is predicted to 
be about 10 $k_{\text{B}}T$ ($k_{\text{B}}$ being the Boltzmann 
constant),\cite{optimumgap} 
one is led to consider small-gap semiconductors
for low-temperature applications. 
If the chemical composition of semiconductors includes transition metals
or rare earths, conduction and valence bands are frequently strongly 
renormalized by correlation effects, forming a temperature-dependent 
gap (see Fig.~\ref{fig2}): this is the case for mixed-valent semiconductors, 
usually cubic, whose relevant electronic properties may be modeled by a 
$f$-flat band and a broad conduction band, with two electrons per unit 
cell.\cite{kondo}
This class of materials consists of two subclasses: The first is the
Kondo insulator, characterized by very strong Coulomb interaction
between electrons on the same rare-earth site, usually 
described by the slave boson solution of the Anderson lattice 
hamiltonian.\cite{slaveboson,quasiferro} 
Mao and Bedell predicted a high value of $ZT$ for
bulk Kondo insulators (the lower the dimension, the higher
the value):\cite{mao} however, some experimental reports
seem to exclude this possibility.\cite{physicstoday,kondom,kondom2}
The second, called the electronic 
ferroelectric (FE),\cite{ferro} consists of semiconductors
with high dielectric constants, such as SmB$_6$ and Sm$_2$Se$_3$,
and it is modeled by the self-consistent mean-field (MF) solution
of the Falicov-Kimball Hamiltonian.\cite{Falicov}
The ground state of the insulating phase is found to be a coherent
condensate of $d$-electron/$f$-hole pairs, giving a net built-in 
macroscopic polarization which breaks the crystal inversion symmetry
and makes the material ferroelectric.\cite{ferro}

Another useful observation is that $ZT$ is reasoned to increase 
in quantum-well superlattices, due to the modification of the density of 
states.\cite{2d,2d2,2d3}
Moreover, superlattices with large thermal impedance
mismatch between layers seem very efficient at
reducing $\kappa_l$ 
because interfaces scatter phonons very effectively.\cite{mahan2d}
However, if the transport is parallel to the layers,
the parasitic $\kappa_l$ of the bigger 
gap material\cite{phonom1,phonom2} or the tunnelling
between conduction layers\cite{phonom2} can drastically decrease
$ZT$. Besides, if intra-layer transport is diffusive 
and described by bulk parameters, $ZT$ for the whole
superlattice cannot be higher than the maximum value for the single
constituents.\cite{toobig}

In addition to the above literature, this work was stimulated by some recent 
advances in thermoelectric applications of junctions. 
Nahum and coworkers\cite{nahum} built an electronic microrefrigerator based
on a metal-insulator-superconductor (NIS) junction.
Subsequent experimental\cite{pekola} and theoretical 
work\cite{edwards,averin,hirsch} confirmed this new idea. Edwards and 
coworkers\cite{edwards} showed that tunneling through structures with 
sharp energy features in the density of states, like quantum dots and
NIS junctions, can be used for cryogenic cooling. 
It seems then a natural extension to us to study the junction between a metal 
and a strongly correlated semiconductor. 

There were recent proposals for devices such as semiconductor/metal 
superlattices with transport perpendicular to interfaces. Mahan and 
Woods\cite{thermionic} suggested a multilayer geometry with the 
thickness of the metallic layer smaller than the electronic mean free 
path and the semiconductor acting as a potential barrier (thermionic 
refrigeration).
Independently, Moyzhes and Nemchinsky\cite{moyzhes} proposed a 
similar configuration with the metallic layer thickness comparable
with the energy relaxation length. We cite also Min and Rowe's idea\cite{min}
of using Fermi-gas / liquid interfaces. 
While the analysis of electronic transport 
of Ref.~\onlinecite{thermionic} and \onlinecite{moyzhes} is not applicable 
to our study, because of the sharp energy profile of the transmission 
coefficient across the junction which we examine, and of the nature of the 
correlated semiconductor ground state, this experimental geometry,
with the strongly correlated semiconductor replacing the barrier layer,
could be implemented, as we suggest at the end of Sec.~\ref{sporcaf}.

In sum, we build on two key ideas from the above literature 
for increasing $ZT$. One is to utilize the sharp energy features in
the density of states of bulk materials as in strongly correlated 
semiconductors. The other is to exploit the good thermoelectric
characteristics of a junction.  In this paper, we combine these ideas 
in exploring the thermopower behavior of a junction between a 
metal and different classes of semiconductors. We describe the gapped material
on one side of the junction as the solution of the Falicov-Kimball 
Hamiltonian\cite{Falicov} in different regimes. In particular, we consider 
the case of: 
(i) Electronic Ferroelectric (FE), where, because of the Coulomb 
interaction between $f$-holes and $d$-electrons, the MF insulating ground 
state is a condensate of excitons;\cite{ferro} 
(ii) Narrow Band semiconductor (NB), characterized
by the $d$-$f$ band hybridization: it would be a Kondo insulator had we take
into account the $f$-$f$ electron repulsion; 
(iii) Broad band semiconductor (SC), for comparison. 
In these three cases, we solve the electronic motion across the junction 
by means of a two-band model analogous to the Bogoliubov-de Gennes 
equations\cite{Degennes} in a finite-difference form. In addition to the 
clean interface, we consider an ``impurity'' overlayer 
made either of rare-earth atoms, with relevant atomic orbitals of $f$-type,
or of atoms with $d$-type orbitals, like transition metals.
In the latter case, electrons can hop from these ``$d$-impurity''
sites to adjacent neighbor atoms, while in the former $f$-impurity case 
hopping is assumed negligible. This scenario is motivated by recent 
advances in atomic layer fabrication. 
We compute $ZT$ for the interface via a linear response.
We find that $ZT$ can be greatly enhanced by the presence
of a suitable $f$-impurity layer at the interface. In particular, 
for a fixed working temperature of the junction,
an optimum energy of the $f$-impurity level exists which maximizes
$ZT$, especially at low temperatures. 
In these regimes, bulk thermal conductivity would be dominated
by phonons which would reduce $ZT$. However, one can fabricate
the junction with two materials with large thermal impedance mismatch, so 
that phonon scatterings at the interface decrease the thermal conductivity. 
Thus, phonon conductivity would not diminish the high $ZT$ found.
To this aim we propose an experimental setup, namely perpendicular 
transport in metal/FE superlattice. 

The structure of the paper is as follows: 
in Section \ref{model} we describe
the model Hamiltonian, in Sec.~\ref{gennes} we solve the
electronic motion across the junction, 
and in Sec.~\ref{trasporto} we compute
transport coefficients and $ZT$.
In Sec.~\ref{risultati} we present
and discuss our results, for the clean interface (\ref{pulita}), and for
the $d$- (\ref{sporcad}) and $f$- 
(\ref{sporcaf}) impurity layer. Also we briefly
discuss some structural relaxation effects of the interface
(sec.~\ref{rilassamento}).
Our conclusions are in Sec.~\ref{fine}.

\section{The model}\label{model}

We introduce the one-dimensional spinless Hamiltonian $\cal{H}$ to model 
the motion across the junction along the $z$ direction perpendicular to
the interface between a metal and different types of semiconductor.
$\cal{H}$ is given by the sum of three terms:
\begin{equation}
{\cal{H}}
={\cal{H}}_{\rm metal}+{\cal{H}}_{\rm interface}+{\cal{H}}_{\rm FK}.
\label{eq:htot}
\end{equation}

${\cal{H}}_{\rm metal}$ is a tight-binding Hamiltonian
describing the metal on the left side of the junction:
\begin{equation}
{\cal{H}}_{\rm metal}=\left(\varepsilon_d^{\prime} + eV\right)\sum_{j<0}
d^{\dagger}_j d_j -t^{\prime}\sum_{j<0} d^{\dagger}_j d_{j+1}
{\rm\; +\; H.c.}
\end{equation}
Here $d_j$ destroys an electron of charge $e<0$
at the lattice site with energy $\varepsilon_d^{\prime}$
and position $z=aj$, $a$ is the lattice constant,
$t^{\prime}$ is the hopping parameter for nearest-neighbor sites, and
$V$ is the electrostatic potential across the junction by applying an external
bias. 
In our model, $V$ is constant in the two bulk regions with a discontinuous
step at the interface, although the actual profile
should be determined self-consistently together with the electronic
density.

\begin{figure}
\centerline{\epsfig{file=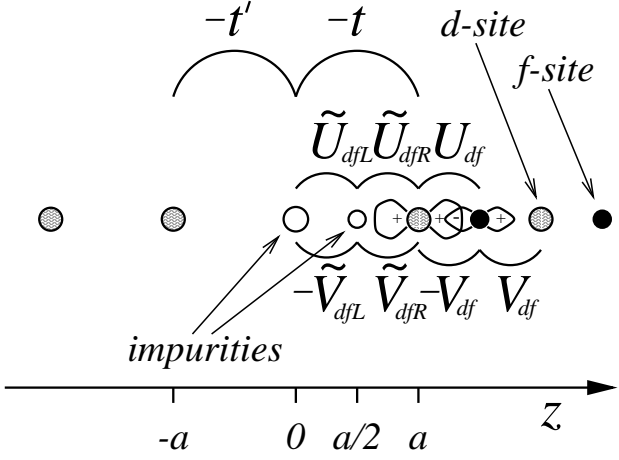,width=6.0cm,,angle=-90}}
\caption{
Pictorial representation of the junction along the $z$-axis perpendicular
to the interface: the atomic sites at $z=aj$ have $d$-type orbitals
while those at $z=a(j+1/2)$ have $f$-type orbitals, even and odd under
spatial inversion, respectively. Different symbols
for the impurities at $z=0$ and $z=a/2$ are drawn; also the parameters
of the Hamiltonian of Eq.~(\ref{eq:htot}) are depicted.
}
\label{fig1}
\end{figure}
The third term ${\cal{H}}_{\rm FK}$ in Eq.~(\ref{eq:htot})
is the Falicov-Kimball Hamiltonian\cite{Falicov} 
referring to the insulator on the right side of the interface.
In addition to the Coulomb interaction $U_{df}$ between the $d$-electrons
at sites $z=aj$ and the $f$-electrons localized on atoms at $z=a(j+1/2)$,
we add a hybridization term ($V_{df}$ is the modulus of the
hybridization integral) between $d$ and $f$ orbitals.
Because the crystal has inversion symmetry, this term has to be odd, 
namely the hybridization term between sites 
at $z=ja$ and $z=\left(j+1/2\right)a$
and that between sites 
at $z=\left(j+1/2\right)a$ and $z=\left(j+1\right)a$
have opposit sign (see Fig.~\ref{fig1}):
\begin{equation}
{\cal{H}}_{\rm FK}={\cal{H}}^{(1)}_{\rm FK}+{\cal{H}}^{(2)}_{\rm FK},
\end{equation}
\begin{eqnarray}
{\cal{H}}^{(1)}_{\rm FK}&=&\tilde{\varepsilon}_d
\sum_{j>0}d^{\dagger}_j d_j
 +\tilde{\varepsilon}_f
\sum_{j>0} f^{\dagger}_{j+1/2} f_{j+1/2}
-t\sum_{j\ge 0} d^{\dagger}_j d_{j+1} 
\nonumber\\  
&-& V_{df}\sum_{j>0} f^{\dagger}_{j+1/2}d_j
+ V_{df}\sum_{j>1}f^{\dagger}_{j-1/2}d_j 
{\rm \;+\; H.c.},
\end{eqnarray}
\begin{eqnarray}
{\cal{H}}^{(2)}_{\rm FK}&=& U_{df}\sum_{j>0}d^{\dagger}_j d_j
f^{\dagger}_{j+1/2} f_{j+1/2} \nonumber\\
&+& U_{df}\sum_{j>1}
d^{\dagger}_j d_j f^{\dagger}_{j-1/2} f_{j-1/2},
\label{eq:two}
\end{eqnarray}
where $t$ is the hopping coefficient, and
$\tilde{\varepsilon}_f$ and $\tilde{\varepsilon}_d$ are the $f$- and
$d$-site energies, respectively.

The second term ${\cal{H}}_{\rm interface}$ in Eq.~(\ref{eq:htot})
is the Hamiltonian at the interface, which describes
the overlayer made of  $d$-sites at $z=0$ and $f$-sites at $z=a/2$:
\begin{eqnarray}
{\cal{H}}_{\rm interface}&=&\tilde{\varepsilon}_{d0}d^{\dagger}_0 d_0+
\tilde{\varepsilon}_{f1/2} f^{\dagger}_{1/2}f_{1/2}\nonumber\\
&+&\tilde{U}_{dfL}d^{\dagger}_0 d_0 f^{\dagger}_{1/2} f_{1/2}
+\tilde{U}_{dfR}d^{\dagger}_1 d_1 f^{\dagger}_{1/2} f_{1/2}\nonumber\\
&&-\tilde{V}_{dfL} f^{\dagger}_{1/2}d_0
+ \tilde{V}_{dfR}f^{\dagger}_{1/2}d_1 
{\rm \;+\; H.c.}
\label{eq:twoint}
\end{eqnarray}
Here we have included the possibility of ``impurity'' atoms at
the interface, namely one with average
energy $\tilde{\varepsilon}_{d0}$ at position $z=0$ and another one
with energy $\tilde{\varepsilon}_{f1/2}$ at position 
$z=a/2$. Since the impurity atoms have orbitals differing from
those in the bulk, the associated $U_{df}$ and $V_{df}$ parameters  
generally change. We denote them by $\tilde{U}_{dfL}$,
$ \tilde{V}_{dfL}$, $\tilde{U}_{dfR}$, and $\tilde{V}_{dfR}$, 
referring to the couples of sites $z=0,a/2$ and $z=a/2,a$, 
respectively (Fig.~\ref{fig1}).

\begin{figure}
\begin{picture}(300,180)
\put(223,166){\epsfig{file=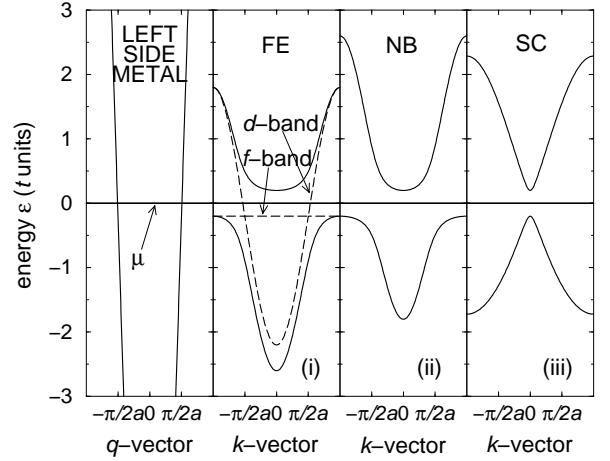,width=2.4in,,angle=-90}}
\end{picture}
\caption{
Quasi-particle bulk energy bands of materials on both sides of
the junction vs wave vector at $T=0$, for the three cases under study,
in the ``semiconductor representation.'' All energies are in units
of $t$, and $\epsilon = 0$ corresponds to the chemical potential $\mu$.
The second panel [case (i)] shows also the ``bare''
band structure (dashed lines), i.e.~that not renormalized by correlation,
with \protect{$V_{df}=\Delta=0$}.
Since the alignment of $\mu$ with respect to the left-hand side metal
band depends on the kind of bulk semiconductor on the right-hand side,
here we assume a metal/FE setup,
with metal (semiconductor) bare bandwidth $4t^{\prime}$ ($4t$),
and $t^{\prime}=5t$. Note that in the case (i)
[electronic ferroelectric] and (ii) [narrow band semiconductor] the
gap is indirect, while it is direct in the case (iii) [broad band
semiconductor].
Besides, in case (i) [(ii)] the bottom (top) of conduction (valence) band
is much flatter than the top (bottom) of valence (conduction) band,
while in case (iii) the curvature
of the two bands close to the direct gap is comparable. The trasformation
\protect{$k\rightarrow \pi/a-k$}, \protect{$\epsilon\rightarrow -\epsilon$}
maps the FE electron (hole) band into the NB hole (electron) band.
In this plot parameters are: (i) $\Delta\!\left(T=0\right)$=0.49$t$,
$V_{df}$=0. (ii) $\Delta$=0, $V_{df}$=0.49$t$. (iii)
$t^{\prime\prime}$=0.141$t$, $V_{\text{e}}$=0.071$t$.
}
\label{fig2}
\end{figure}

We give the MF solution of the
Hamiltonian $\cal{H}$ of Eq.~(\ref{eq:htot}), by means of 
an approach (see Sec.~\ref{gennes}) analogous
to the Bogoliubov-de Gennes method\cite{Degennes} (BdG).  
In contrast to the electron-electron pairing in
the Bardeen-Cooper-Schrieffer (BCS) MF theory of 
superconductivity, the pairing here
occurs between $d$-electrons and $f$-holes. To proceed, we assume that
\begin{equation}
 \left<d^{\dagger}_j d_j\right>=n_d \qquad \forall j\ge 1,
\label{eq:ansa1}
\end{equation}
\begin{equation}
\left<f^{\dagger}_{j+1/2} f_{j+1/2}\right>=n_f \qquad
\forall j\ge 1,
\end{equation}
\begin{eqnarray}
U_{df}\left<d^{\dagger}_j f_{j+1/2}\right>&=&\Delta\qquad \forall j\ge 1,
\nonumber\\
U_{df}\left<d^{\dagger}_j f_{j-1/2}\right>&=&\Delta \qquad 
\forall j\ge 2,
\label{eq:deltadefinit}
\end{eqnarray}
\begin{equation}
 \left<d^{\dagger}_0 d_0\right>=n_{d0},\qquad
\left<f^{\dagger}_{1/2} f_{1/2}\right>=n_{f1/2},
\end{equation}
\begin{equation}
\tilde{U}_{dfL}\left<d^{\dagger}_0 f_{1/2}\right>=\tilde{\Delta}_L,
\end{equation}
\begin{equation}
\tilde{U}_{dfR}\left<d^{\dagger}_1 f_{1/2}\right>=\tilde{\Delta}_R.
\label{eq:ansan}
\end{equation}
Here $\left<\ldots\right>$ is the symbol for the quantum statistical average.
Note that, in addition to the usual mean orbital occupations $n_d$ and
$n_f$ of standard Hartree-Fock theory ($0\leq n_d, n_f \leq1$), 
we also introduce 
the non-vanishing pairing potential $\Delta$, 
characteristic built-in coherence of
the $d$-electron/$f$-hole condensate.
Following de Gennes,\cite{Degennes} we now write an effective Hamiltonian 
$\cal{H}^{\rm MF}$, to be computed self-consistently 
together with the energy spectrum:
\begin{equation}
{\cal{H}}^{\rm MF}={\cal{H}}_{\rm metal}+{\cal{H}}_{\rm interface}^{\rm MF}+
{\cal{H}}_{\rm FK}^{\rm MF},
\label{eq:hscf}
\end{equation}
\begin{eqnarray}
{\cal{H}}_{\rm interface}^{\rm MF}&=&
\varepsilon_{d0}d^{\dagger}_0 d_0+
\varepsilon_{f1/2} f^{\dagger}_{1/2}f_{1/2}
-\tilde{\Delta}_{L} f^{\dagger}_{1/2} d_0 \nonumber\\
&-&\tilde{\Delta}_{R} f^{\dagger}_{1/2} d_1 
- \tilde{V}_{dfL} f^{\dagger}_{1/2}d_0
+ \tilde{V}_{dfR}f^{\dagger}_{1/2}d_1 
{\rm \;+\; H.c.} \nonumber\\
&&-\tilde{U}_{dfL}n_{d0}n_{f1/2} +\left|\tilde{\Delta}_{L}\right|^2
+\left|\tilde{\Delta}_{R}\right|^2,
\label{eq:mf1}
\end{eqnarray}
\begin{eqnarray}
{\cal{H}}_{\rm bulk}^{\rm MF} &=& 
\varepsilon_{d1}d^{\dagger}_1 d_1+
\varepsilon_d
\sum_{j>1}d^{\dagger}_j d_j+ \varepsilon_f
\sum_{j>0} f^{\dagger}_{j+1/2} f_{j+1/2} \nonumber\\
&-& t\sum_{j\ge 0} d^{\dagger}_j d_{j+1} 
-V_{df}\sum_{j>0} f^{\dagger}_{j+1/2}d_j \nonumber\\
&+& V_{df}\sum_{j>1}f^{\dagger}_{j-1/2}d_j
-\Delta \sum_{j>0}f^{\dagger}_{j+1/2} d_j \nonumber\\
&-& \Delta \sum_{j>1} f^{\dagger}_{j-1/2} d_j
{\rm \;+\; H.c.} -\tilde{U}_{dfR}n_d n_{f1/2} \nonumber\\
&-& U_{df}n_d n_f-\sum_{j>1}2U_{df}n_d n_f
+\sum_{j>0}2\left|\Delta\right|^2.
\label{eq:mf2}
\end{eqnarray}
Here we have defined the renormalized energies
\begin{eqnarray}
\varepsilon_{d0} &=& \tilde{\varepsilon}_{d0}+\tilde{U}_{dfL}\, n_{f1/2},
\nonumber\\
\varepsilon_{f1/2} &=& \tilde{\varepsilon}_{f1/2}+
\tilde{U}_{dfL} \,n_{d0}+\tilde{U}_{dfR}\, n_d,\nonumber\\
\varepsilon_{d1} &=& \tilde{\varepsilon}_d+\tilde{U}_{dfR}\, n_{f1/2}
+U_{df}\,n_f, \nonumber\\  
\varepsilon_f &=& \tilde{\varepsilon}_f
+2 U_{df} \,n_d,\nonumber\\
\varepsilon_d&=&\tilde{\varepsilon}_d+2 U_{df} \,n_f.
\end{eqnarray}
We shall treat the renormalized quantities $\varepsilon_{d0}$,
$\varepsilon_{f1/2}$, $\varepsilon_{d1}$, $\varepsilon_f$, and
$\varepsilon_d$ as material parameters. For 
simplicity, we shall assume $t^{\prime}>t$, $\varepsilon_{d1}=\varepsilon_d$,
and $\varepsilon_d^{\prime}=\varepsilon_d=0$, that is the middle of
the $d$-band on both sides of the junction is aligned, 
but the metal bandwidth ($4t^{\prime}$) is larger than the 
semiconductor bandwidth (\protect{$4t$}).
We will consider only the case $\varepsilon_f=0$, that is the flat band
is in the middle of the $d$-band in the right-side material
(see Fig.~\ref{fig2}).\cite{czycholl}
With these assumptions, and dropping all the constant terms in 
Eqs.~(\ref{eq:mf1}-\ref{eq:mf2}), we obtain the expression of the 
Hamiltonian ${\cal{H}}^{\rm MF}$  which we will use in the following:
\begin{eqnarray}
{\cal{H}}^{\rm MF}&=&
eV\sum_{j<0} d^{\dagger}_j d_j 
+\varepsilon_{d0}d^{\dagger}_0 d_0+
\varepsilon_{f1/2} f^{\dagger}_{1/2}f_{1/2}\nonumber\\
&-& t^{\prime}\sum_{j<0} d^{\dagger}_j d_{j+1}
-t\sum_{j\ge 0} d^{\dagger}_j d_{j+1}   \nonumber\\
&& -( \tilde{\Delta}_{L}+\tilde{V}_{dfL}) f^{\dagger}_{1/2} d_0
-(\tilde{\Delta}_{R}-\tilde{V}_{dfR}) f^{\dagger}_{1/2} d_1 
\nonumber\\
&& \quad -(\Delta+V_{df}) \sum_{j>0}f^{\dagger}_{j+1/2} d_j \nonumber\\
&& \qquad -(\Delta-V_{df})\sum_{j>1}f^{\dagger}_{j-1/2} d_j
{\rm \;+\; H.c.}
\label{eq:h}
\end{eqnarray}
Note the different parity of $\Delta$- and $V_{df}$-terms: 
while the $V_{df}$-term is odd, the $\Delta$-term is even.\cite{quasiferro}

The values of averages on the left side of 
equations (\ref{eq:ansa1}-\ref{eq:ansan}) are spatially 
inhomogeneous and should be determined self-consistently 
and simultaneously with the solution
of the BdG equations. We make the approximation that those are the 
bulk averages appropriate to each side of the interface layer. 
In particular, we take the order parameter $\Delta$ as constant
on the right side of the interface, and zero on the left 
side.\cite{interface} The quantities
$\tilde{\Delta}_L$, $\tilde{\Delta}_{R}$ are used to characterize the effect
of the interface on $\Delta$.

The temperature dependence of $\Delta$ is much stronger than that of the
average occupation of orbitals. We determine numerically $\Delta$,
together with the chemical potential $\mu$, as the bulk value deep inside 
the right material in a self-consistent way. For this purpose we use the 
bulk Hamiltonian 
${\cal{H}}^{\rm MF}_{\rm bulk}$:
\begin{eqnarray}
{\cal{H}}^{\rm MF}_{\rm bulk} &=&
-t\sum_jd^{\dagger}_j d_{j+1}-\left(\Delta+V_{df}\right) 
\sum_jf^{\dagger}_{j+1/2} d_j \nonumber\\
&& -\left(\Delta-V_{df}\right)\sum_jf^{\dagger}_{j-1/2} d_j
{\rm \;+\; H.c.}
\label{eq:inside}
\end{eqnarray}
Here the index $j$ runs over the whole space. We always assume an 
electronic occupation of one electron per unit cell (a cell contains one 
$d$-site and one $f$-site), namely
\begin{equation}
\frac{1}{N_s}\sum_j\left<d^{\dagger}_j d_j + f^{\dagger}_{j+1/2}
f_{j+1/2}\right>=1,
\label{eq:numb}
\end{equation}
where $N_s$ is the number of cells. Eq.~(\ref{eq:numb}) allows us to calculate
the chemical potential $\mu$ for the bulk. Because the effective Hamiltonian
of Eq.~(\ref{eq:inside}) is strongly temperature-dependent, so $\mu$ is. 
Since $\mu$ refers to the bulk value inside the semiconductor, 
the metal on the left side acts as an electron reservoir.

We will consider three specific cases for the actual values of
parameters in the Hamiltonian ${\cal{H}}^{\rm MF}$ of Eq.~(\ref{eq:h}):
(i) $V_{df}\rightarrow 0$ and non-zero $U_{df}$ 
($\tilde{V}_{dfL}=\tilde{V}_{dfR}=0$), i.e.~the case FE
of the electronic ferroelectric with negligible 
hybridization.\cite{gaugeinvariance}
(ii) $U_{df}\rightarrow 0$ and non-zero $V_{df}$ 
($\tilde{U}_{dfL}=\tilde{U}_{dfR}=0$), i.e.~the case NB of a $f$-band 
hybridized with a $d$-band, that is a narrow band
semiconductor which would be a Kondo insulator if it were driven by
the condensation of slave bosons.
(iii) A ordinary, ``broad band'' semiconductor SC, without center of
inversion. In the latter case we still use ${\cal{H}}^{\rm MF}$ but we 
regard it simply as a one particle Hamiltonian where the ``$d$'' and ``$f$''
indices are pure labels, and we rename $V_{df}$ as $t$, considering it as
the odd part of a hopping coefficient between nearest neighbor sites, 
$\Delta$ as $V_{\text{e}}$, the even part, and $t$ as $t^{\prime\prime}$, 
a second nearest neighbor hopping coefficient. We make the choice
$t\gg V_{\text{e}}$, $V_{\text{e}}\sim t^{\prime\prime}$, leading to broad 
conduction and valence bands with different effective masses.

While in case (i) $\Delta$ has to be determined self-consistently 
and in case (ii) $V_{df}$ is a material parameter, there is
a formal canonical transformation connecting (i) to (ii),
\begin{eqnarray}
d_j\rightarrow d_j^{\dagger}{\rm e}^{-{\rm i}\pi j},
&\qquad&f_{j+1/2}\rightarrow f_{j+1/2}^{\dagger}{\rm e}^{-{\rm i}\pi (j+1)},
\nonumber\\
\Delta\rightarrow V_{df}^*,&\qquad&\varepsilon_{d0}\rightarrow-
\varepsilon_{d0},\nonumber\\
\varepsilon_{f1/2}\rightarrow-\varepsilon_{f1/2},&\qquad&
eV\rightarrow -eV,\nonumber\\
\tilde{\Delta}_L\rightarrow \tilde{V}_{dfL}^*,&\qquad&
\tilde{\Delta}_R\rightarrow \tilde{V}_{dfR}^*,
\label{eq:canonical}
\end{eqnarray}
leading to the mapping:
\begin{eqnarray}
{\cal{H}}^{\rm MF}\!\left(V_{df}=0\right)
& \rightarrow & {\cal{H}}^{\rm MF}\!\left(\Delta=0\right) \nonumber\\
&& -\sum_{j<0}eV-\varepsilon_{d0}-\varepsilon_{f1/2}.
\end{eqnarray}
This transformation, in particular, maps the FE electron 
(hole) excitations 
into the NB hole (electron) excitations.

\section{Transport across the junction}\label{formalism}

In the first subsection, we solve the BdG equations; in the second one,
we compute the transport coefficients, in the limit of small electric or 
thermal gradient across the junction.

\subsection{The Bogoliubov-de Gennes equations}\label{gennes}

We now find the canonical transformation that diagonalizes 
${\cal{H}}^{\rm MF}$ in a self-consistent way, i.e., we solve the BdG 
equations.  In the present context, the BdG approach is essentialy the
Hartree-Fock treatment of the two-band model.  It is convenient to 
refer all the excitation energies to the chemical potential $\mu$. Thus, 
we define the number operator $N$,
\begin{displaymath}
N=\sum_j\left(d^{\dagger}_j d_j + f^{\dagger}_{j+1/2} f_{j+1/2}\right),
\end{displaymath}
and we replace the Hamiltonian ${\cal{H}}^{\rm MF}$ with
${\cal{H}}^{\rm MF}\!-\mu N$.
To diagonalize it, we introduce the unitary transformation
\begin{eqnarray}
\gamma_{ke}&=&\frac{1}{\sqrt{N_s}}  \sum_j   [
u_k^*\!\left(j\right)d_j + v_k^*\!\left(j+1/2\right)f_{j+1/2} ]
, \nonumber\\
\gamma^{\dagger}_{-kh} &=& \frac{1}{\sqrt{N_s}}  \sum_j  [
\bar{u}_{-k}\!\left(j\right)d_j \nonumber\\
&&  \qquad + \quad \bar{v}_{-k}
\!\left(j+1/2\right)f_{j+1/2}],
\label{eq:direct}
\end{eqnarray}
where $k$ is  simply a quantum index, equivalent to the crystal momentum 
only in the bulk case.  The idea is that the operators 
$\gamma_{ke}^{\dagger}$, $\gamma_{-kh}^{\dagger}$ must diagonalize 
${\cal{H}}^{\rm MF}\!-\mu N$ and create
elementary quasi-particle excitations of energy $\omega\!\left(k\right)$
(electrons) and $\bar{\omega}\!\left(-k\right)$ (holes), respectively,
if applied to the ground state.
Therefore the equations of motion for the operators $\gamma_{ke}$ 
and $\gamma_{-kh}$ are
\begin{eqnarray}
{\rm i}\hbar\dot{\gamma}_{ke}&=&\left[\gamma_{ke},{\cal{H}}^{\rm MF}
\!-\mu N\right]=
\omega\!\left(k\right)\gamma_{ke},\nonumber\\
{\rm i}\hbar\dot{\gamma}_{-kh}&=&\left[\gamma_{-kh},{\cal{H}}^{\rm MF}
\!-\mu N\right]=
\bar{\omega}\!\left(-k\right)\gamma_{-kh},
\label{eq:key}
\end{eqnarray}
and the Hamiltonian
${\cal{H}}^{\rm MF}\!-\mu N$ acquires the form
\begin{eqnarray}
{\cal{H}}^{\rm MF}\!-\mu N &=& \sum_k\;\Big[\,
\omega\!\left(k\right)\gamma_{ke}^{\dagger}
\gamma_{ke} \nonumber\\
&& + \quad \bar{\omega}\!\left(-k\right)
\gamma_{-kh}^{\dagger}\gamma_{-kh}
\Big].
\label{eq:fermgas}
\end{eqnarray}
Note that in our definition, Eqs.~(\ref{eq:key}), the operator
$\gamma_{-kh}^{\dagger}$ is a fermionic creation operator
which {\em excites} a hole, hence the hole energy,
as well as the electron energy, is
positive, i.e.~$\omega\!\left(k\right)>0$, 
$\bar{\omega}\!\left(-k\right)>0$ [``excitation representation'' (ER)].
Besides, $\omega\!\left(k\right)$, $\bar{\omega}\!\left(-k\right)$
depend on the chemical potential $\mu$ and on the built-in coherence
$\Delta$, and consequently on the temperature $T$.
The $\gamma$ operators are analogous to the Bogoliubov-Valatin operators 
of the BCS theory; however, in the present context the 
eigenstates of the hamiltonian
${\cal{H}}^{\rm MF}$ have definite numbers of particles, 
namely $\gamma_{ke}^{\dagger}$
creates an electron and $\gamma_{-kh}^{\dagger}$ creates a hole, while 
the particle number of the BCS solution is an average quantity,
and in that case $\gamma^{\dagger}$ creates a quasi-particle which is a
mixture of electron and hole.
\begin{widetext}
From Eqs.~(\ref{eq:key})
and the condition that
the coefficients of each independent 
$\gamma$-operator must be zero, we obtain finite-difference 
equations for the electron site-coefficients
$u$ and $v$ of the junction.
For the bulk semiconductor the BdG equations are:
\begin{eqnarray}
&& \left[\omega\!\left(k\right)+\mu\right]u_k\!\left(j\right) =
-tu_k\!\left(j-1\right)-tu_k\!\left(j+1\right)
-\left(\Delta^*-V_{df}^*\right)v_k\!\left(j-1/2\right)
-\left(\Delta^*+V_{df}^*\right)v_k\!\left(j+1/2\right)\qquad
\forall j>1,\nonumber\\
&& \left[\omega\!\left(k\right)+\mu\right]v_k\!\left(j+1/2\right) =
-\left(\Delta+V_{df}\right)u_k\!\left(j\right)
-\left(\Delta-V_{df}\right)u_k\!\left(j+1\right)\qquad \forall j>0.
\label{eq:BdGstar}
\end{eqnarray}
Similarly, the hole amplitudes $\bar{u}$ and $\bar{v}$ are
given by:
\begin{eqnarray}
&&
\left[\bar{\omega}\!\left(-k\right)-\mu\right]\bar{u}_{-k}\!\left(j\right)=
t\bar{u}_{-k}\!\left(j-1\right)+t\bar{u}_{-k}\!\left(j+1\right)
+\left(\Delta-V_{df}\right)\bar{v}_{-k}\!\left(j-1/2\right)
+\left(\Delta+V_{df}\right)\bar{v}_{-k}\!\left(j+1/2\right)
\qquad \forall j>1,\nonumber\\
&& \left[\bar{\omega}\!\left(-k\right)-\mu\right]
\bar{v}_{-k}\!\left(j+1/2\right) =
\left(\Delta^*+V_{df}^*\right)\bar{u}_{-k}\!\left(j\right)
+\left(\Delta^*-V_{df}^*\right)\bar{u}_{-k}\!\left(j+1\right)
\qquad \forall j>0.
\label{eq:BdGend}
\end{eqnarray}
\end{widetext}

To solve Eq.~(\ref{eq:BdGstar}), we write down the
trial two-component wavefunction
\begin{equation}
{u_k\!\left(j\right) \choose v_k\!\left(j+1/2\right)}=
{u_k \choose v_k{\rm e}^{{\rm i}ka/2}}
{\rm e}^{{\rm i}kaj}.
\label{eq:provap}
\end{equation}
In the following, we take  
without loss of generality $u_k>0$, and $\Delta$, $V_{df}$ 
($\tilde{\Delta}_L$, $\tilde{\Delta}_R$, ${\tilde{V}}_{dfL}$,
${\tilde{V}}_{dfR}$) real.
Equation (\ref{eq:provap}) is compatible with (\ref{eq:BdGstar}) only if
\begin{equation}
\omega\!\left(k\right)=\xi_k\pm E_k -\mu,
\label{eq:omegap}
\end{equation}
where
\begin{displaymath}
\xi_k=\varepsilon_k/2,\qquad\qquad\varepsilon_k=
-2t\cos{(ka)},
\end{displaymath}
\begin{displaymath}
E_k=\sqrt{\xi_k^2+\left|\Delta_k-V_k\right|^2},
\end{displaymath}
\begin{equation}
\Delta_k=2\Delta\cos{(ka/2)},\qquad 
V_k=2{\rm i}V_{df}\sin{(ka/2)}.
\end{equation}
Since $\omega>0$, we choose the sign $+$ in Eq.~(\ref{eq:omegap}). 
Consequently,
the amplitudes $\left(u_k,v_k\right)$ are given by
\begin{equation}
\left(V_k-\Delta_k\right)u_k=\left(\xi_k+E_k\right)v_k,
\label{eq:phase}
\end{equation}
plus the normalization condition 
\begin{equation}
u_k^2+\left|v_k\right|^2=1,
\end{equation} 
that is 
\begin{equation}
u_k^2=\frac{1}{2}\left(1+\frac{\xi_k}{E_k}\right),\qquad
\left|v_k\right|^2=\frac{1}{2}\left(1-\frac{\xi_k}{E_k}\right),
\end{equation}
and the phase of $v_k$ is determined by Eq.~(\ref{eq:phase}).
In the same way, the solution for holes is (\protect{$\bar{u}_{-k}>0$})
\begin{equation}
{\bar{u}_{-k}\!\left(j\right) \choose \bar{v}_{-k}\!\left(j+1/2\right)}=
{\bar{u}_{-k} \choose \bar{v}_{-k}{\rm e}^{-{\rm i}ka/2}}
{\rm e}^{-{\rm i}kaj},
\label{eq:provam}
\end{equation}
with energy
\begin{equation}
\bar{\omega}\!\left(-k\right)=-\xi_{-k}+ \bar{E}_{-k} +\mu,
\label{eq:omegam}
\end{equation}
where
\begin{equation}
\bar{E}_{-k}=
\sqrt{\xi_{-k}^2+\left|\Delta_{-k}+V_{-k}\right|^2},
\end{equation}
and amplitudes
\begin{eqnarray}
\bar{u}_{-k}^2 &=& \frac{1}{2}\left(1-\frac{\xi_{-k}}{\bar{E}_{-k}}\right),
\nonumber\\
\left|\bar{v}_{-k}\right|^2 &=& \frac{1}{2}\left(1+
\frac{\xi_{-k}}{\bar{E}_{-k}}\right),
\end{eqnarray}
and the phase of $\bar{v}_{-k}$ determined by the equation
\begin{equation}
\left(\xi_{-k}+\bar{E}_{-k}\right)\bar{u}_{-k}=
\left(\Delta_{-k}+V_{-k}\right)\bar{v}_{-k}.
\end{equation}
Figure \ref{fig2} shows typical quasi-particle band structures
on both sides of the junction 
for the three different cases under study. In this picture the
energy branches are drawn according to the ``semiconductor
representation'' (SCR), where the 
quasi-particle energy $\epsilon$ for holes is negative, namely
\begin{displaymath}
\epsilon=\cases{\omega & for electrons \cr
-\bar{\omega} & for holes. \cr}
\end{displaymath}
In SCR the ground state (vacuum of $\gamma$-operators in ER) 
can be represented regarding 
the valence band as all filled with electrons and the conduction
band empty.
Note that the asymmetry of conduction and valence bands 
close to the gap is remarkable 
in case (i) and (ii), contrary to case (iii). 
For FE (NB) the gap is indirect, and the bottom (top) of
conduction (valence) band is much flatter than the top (bottom) of
valence (conduction) band,
while the curvature of the two SC bands close to the direct gap is
comparable. 
Besides, as a consequence of the transformation
(\ref{eq:canonical}), the mapping
\protect{$k\rightarrow \pi/a-k$}, \protect{$\epsilon\rightarrow -\epsilon$}
transforms the FE electron (hole) band into the NB hole (electron) band. 
 
Once we know the semiconductor bulk energy spectrum, we can also compute the
probability current density \protect{$J_{keN}\!\left(j\right)$}
[\protect{$J_{-khN}\!\left(j\right)$}] associated
with the quasi-particle wavefunction \protect{$\left(u,v\right)$}
[\protect{$\left(\bar{u},\bar{v}\right)$}] (see appendix \ref{appcurrent}).
From Eq.~(\ref{eq:enumber}-\ref{eq:hcurrentbulk}), 
(\ref{eq:provap}) and
(\ref{eq:provam}) 
we obtain
\begin{eqnarray}
J_{keN}\!\left(j\right) &=&
\frac{2t}{\hbar}u_k^2\sin{\left(ka\right)}
+\frac{1}{\hbar}u_k\left|v_k\right| 
\Big[\left(\Delta+V_{df}\right) \nonumber\\
&\times& \sin{\left(ka/2+\varphi\right)}
+ \left(\Delta-V_{df}\right) \nonumber\\
&\times & \sin{\left(ka/2-\varphi\right)}\Big],\qquad v_k=\left|v_k\right|
{\rm e}^{{\rm i}\varphi},
\end{eqnarray}
\begin{eqnarray}
J_{-khN}\!\left(j\right)&=&\frac{2t}{\hbar}\bar{u}_{-k}^2
\sin{\left(ka\right)}
+\frac{1}{\hbar}\bar{u}_{-k}\left|\bar{v}_{-k}\right|
\Big[\left(\Delta+V_{df}\right) \nonumber\\
&\times & \sin{\left(ka/2-\bar{\varphi}\right)}
+\left(\Delta-V_{df}\right) \nonumber\\
&\times & \sin{\left(ka/2+\bar{\varphi}\right)}\Big],
\qquad \bar{v}_{-k}=\left|\bar{v}_{-k}\right|
{\rm e}^{{\rm i}\bar{\varphi}}.
\end{eqnarray}
Clearly the bulk current is site-independent. It can be shown that
\begin{equation}
J_{keN}=\frac{1}{a\hbar}
\frac{\partial\, \omega\!\left(k\right)}{\partial k},
\qquad\qquad J_{-khN}=\frac{1}{a\hbar}
\frac{\partial\, \bar{\omega}\!\left(-k\right)}{\partial \left(-k\right)},
\label{eq:semiclassic}
\end{equation}
i.e.~the quasi-particle velocity has the same expression as the
semi-classical one.
According to formulae (\ref{eq:semiclassic}) 
for the probability density current,
the solution (\ref{eq:provap}) [(\ref{eq:provam})]
represents a two-component electron (hole)
wavefunction with wave vector $k$ ($-k$) travelling from left to right
if $k>0$.
In the bulk we assume periodic
boundary conditions for $k$ and restrict its values to the
first Brillouin zone, namely 
\protect{$k=(2\pi/a)\,n/N_s$}, \protect{$-N_s/2<n\leq N_s/2$}, $N_s$ even.

Let us  focus on case (i). In order to compute the temperature 
dependence of the order parameter $\Delta$, recall definition 
(\ref{eq:deltadefinit}) and use the inverse transformation of 
(\ref{eq:direct}) to obtain 
\begin{eqnarray}
\Delta &=& \frac{U_{df}}{N_s}\sum_k\Big\{\bar{u}_{-k}\!
\left(j\right)\bar{v}_{-k}^*\!\left(j+1/2\right)\left<\gamma_{-k h}
\gamma_{-k h}^{\dagger} \right> \nonumber\\
&&+ \quad
u^*_k\!\left(j\right)v_k\!\left(j+1/2\right)
\left<\gamma_{k e}^{\dagger}\gamma_{k e}\right> \Big\}.
\end{eqnarray}
The average occupations $\left<\gamma_{ke}^{\dagger}\gamma_{ke}\right>$,
$\left<\gamma_{-kh}^{\dagger}\gamma_{-kh}\right>$ are simply given by the
Fermi function $f\!\left(\omega\right)$
\begin{equation}
\left<\gamma_{ke}^{\dagger}\gamma_{ke}\right>=
f\!\left(\omega\!\left(k\right)\right), \quad
\left<\gamma_{-kh}^{\dagger}\gamma_{-kh}\right>=
f\!\left(\bar{\omega}\!\left(-k\right)\right),
\label{eq:mean}
\end{equation}
with
\begin{displaymath}
f\!\left(\omega\right)=\left({\rm e}^{\beta\omega}+1\right)^{-1}
\qquad\qquad\qquad \beta=1/k_{\text{B}} T.
\end{displaymath}
Replacing the $(u,v)$ amplitudes with the values we have just computed,
Eq.~(\ref{eq:mean}) turns into
\begin{eqnarray}
\Delta\!\left(T\right) &=& \frac{U_{df}}{N_s}\sum_k\Big\{
\frac{\Delta_k}{2E_k}\cos{\left(ka/2\right)} \nonumber\\
&& \times\quad \Big[1-
f\!\left(\bar{\omega}\!\left(-k\right)\right)
-f\!\left(\omega\!\left(k\right)\right)\Big]\Big\},
\label{eq:BCSgapbis}
\end{eqnarray}
which is the analogous of the BCS gap equation,
implicitely defining $\Delta\!\left(T\right)$. 
The restraint (\ref{eq:numb})
on the electron number is an implicit definition of $\mu\!
\left(T\right)$; in terms of $\gamma$ operators it turns into:
\begin{equation}
\frac{1}{N_s}\sum_k f\!\left(\omega\!\left(k\right)\right)=
\frac{1}{N_s}\sum_k f\!\left(\bar{\omega}\!\left(-k\right)\right).
\label{eq:mu}
\end{equation}
We solve both Eqs.~(\ref{eq:BCSgapbis}) and (\ref{eq:mu}) simultaneously
to obtain the value of $\Delta\!\left(T\right)$ and $\mu\!\left(T\right)$
deep inside the FE bulk. 
A critical temperature $T_C$ exists beyond which \protect{$\Delta=\mu
=0$} and the energy gap of the material vanishes.

Since we know the bulk solution on the semiconductor side of the
junction, we now study the motion of quasi-particles along the 
whole junction.
The BdG equations for the bulk of the metal on the left side
\begin{eqnarray}
&& \left[\omega\!\left(k\right)+\mu\right]u_k\!\left(j\right)=
eVu_k\!\left(j\right) \nonumber\\
&&\quad -t^{\prime}u_k\!\left(j-1\right)
-t^{\prime}u_k\!\left(j+1\right),\nonumber\\
&& \left[\omega\!\left(k\right)+\mu\right]v_k\!\left(j+1/2\right)=
0 \quad \forall j<0,
\label{eq:BdGbisstar}
\end{eqnarray}
\begin{eqnarray}
&& \left[\bar{\omega}\!\left(-k\right)-\mu\right]\bar{u}_{-k}\!\left(j\right)=
-eV\bar{u}_{-k}\!\left(j\right) \nonumber\\
&& \quad +t^{\prime}\bar{u}_{-k}\!\left(j-1\right)
+t^{\prime}\bar{u}_{-k}\!\left(j+1\right),\nonumber\\
&& \left[\bar{\omega}\!\left(-k\right)-\mu\right]\bar{v}_{-k}
\!\left(j+1/2\right)=
0\quad \forall j<0,
\label{eq:BdGbisend}
\end{eqnarray}
have the Bloch solution
\begin{eqnarray}
{u_k\!\left(j\right) \choose v_k\!\left(j+1/2\right)} &=&
{1 \choose 0}{\rm e}^{{\rm i}qaj},\nonumber\\
{\bar{u}_{-k}\!\left(j\right) \choose \bar{v}_{-k}\!\left(j+1/2\right)} &=&
{1 \choose 0}{\rm e}^{-{\rm i}qaj},
\end{eqnarray}
with energy
\begin{eqnarray}
\omega^{\prime}\!\left(q\right) &=& eV-2t^{\prime}\cos{\left(qa\right)}-\mu,
\nonumber\\
\bar{\omega}^{\prime}\!\left(-q\right) &=&
-eV+2t^{\prime}\cos{\left(-qa\right)} +\mu,
\label{eq:energyl}
\end{eqnarray}
and, from Eq.~(\ref{eq:je}-\ref{eq:jh}), probability current density
\begin{eqnarray}
J_{keN}\!\left(j\right) &=&
\frac{2t^{\prime}}{\hbar}\sin{\left(qa\right)},
\nonumber\\ J_{-khN}\!\left(j\right) &=&
-\frac{2t^{\prime}}{\hbar}\sin{\left(-qa\right)},
\end{eqnarray}
or, equivalently,
\begin{equation}
J_{keN}=\frac{1}{a\hbar}
\frac{\partial\, \omega^{\prime}\!\left(q\right)}{\partial q},
\quad J_{-khN}=\frac{1}{a\hbar}
\frac{\partial\, \bar{\omega}^{\prime}\!
\left(-q\right)}{\partial \left(-q\right)}.
\label{eq:semiclassic2}
\end{equation}

The idea is to match in some way the bulk solutions on both sides of
the junction.
To make the discussion easier, let us
consider only the electron motion.
The physical boundary condition is that the electron travels e.g.~from 
$z=-\infty$ towards positive values of $z$ and it is partly 
transmitted through the junction
and partly reflected. Note that the same energy $\omega$ corresponds
to two different bulk wave vectors $q$ and $k$, 
while in metal/superconductor junctions
the wave vector can be approximated by the Fermi vector on
both sides of the interface.
We always use $k$ to label the coherent electronic 
state through the whole space, with the convention that both
$k$ and $q$, wave vectors in their respective bulks, correspond to the
same energy $\omega$.  
It is convenient to define the wavefunctions
\begin{equation}
{\Psi_{1L}\!\left(j\right) \choose \Psi_{2L}\!\left(j+1/2\right)}
={1 \choose 0}{\rm e}^{{\rm i}qaj}-R_k{1 \choose 0}{\rm e}^{-{\rm i}qaj}
\quad \forall j,
\label{eq:incoming}
\end{equation}
\begin{equation}
{\Psi_{1R}\!\left(j\right) \choose \Psi_{2R}\!\left(j+1/2\right)}
=T_k{u_k \choose v_k{\rm e}^{{\rm i}ka/2}}{\rm e}^{{\rm i}kaj}
\quad \forall j,
\label{eq:transmitted}
\end{equation}
with the elastic scattering condition
\begin{equation}
\omega^{\prime}\!\left(q\right)=\omega\!\left(k\right).
\end{equation}
If we define
\begin{displaymath}
{u_k\!\left(j\right) \choose v_k\!\left(j+1/2\right)}
\end{displaymath}
as the solution of the motion across the whole junction
$\forall j$, we immediately have from
Eq.~(\ref{eq:BdGbisstar}) that
\begin{eqnarray}
{u_k\!\left(j\right) \choose v_k\!\left(j+1/2\right)} &=&
{\Psi_{1L}\!\left(j\right) \choose \Psi_{2L}\!\left(j+1/2\right)}
\quad\forall j<0, \nonumber\\
u_k\!\left(0\right) &=& \Psi_{1L}\left(0\right),
\end{eqnarray}
and from Eq.~(\ref{eq:BdGstar}) that
\begin{equation} 
{u_k\!\left(j\right) \choose v_k\!\left(j+1/2\right)}=
{\Psi_{1R}\!\left(j\right) \choose \Psi_{2R}\!\left(j+1/2\right)}
\quad\forall j>0,
\end{equation}
because (\ref{eq:BdGstar}) and (\ref{eq:BdGbisstar})
are linear and homogeneous. 
We have still to determine
$v_k\!\left(1/2\right)$ and the two unknown constants $T_k$ and $R_k$, so
we need the three BdG equations at the interface not yet employed:
\begin{widetext}
\begin{equation}
\left[\omega\!\left(k\right)+\mu\right]u_k\!\left(0\right)=
\varepsilon_{d0}u_k\!\left(0\right)
-t^{\prime}u_k\!\left(-1\right)-tu_k\!\left(1\right)
-\left(\tilde{\Delta}_L^*+\tilde{V}_{dfL}^*\right)
v_k\!\left(1/2\right),
\label{eq:BdGe}
\end{equation}
\begin{equation}
\left[\omega\!\left(k\right)+\mu\right]v_k\!\left(1/2\right)=
\varepsilon_{f1/2}v_k\!\left(1/2\right)-
\left(\tilde{\Delta}_L+\tilde{V}_{dfL}\right)u_k\!\left(0\right)
-\left(\tilde{\Delta}_R-\tilde{V}_{dfR}\right)u_k\!\left(1\right),
\label{eq:BdGc}
\end{equation}
\begin{equation}
\left[\omega\!\left(k\right)+\mu\right]u_k\!\left(1\right)=
-tu_k\!\left(0\right)-tu_k\!\left(2\right)
-\left(\tilde{\Delta}_R^*-\tilde{V}_{dfR}^*\right)v_k\!\left(1/2\right)
-\left(\Delta^*+V_{df}^*\right)v_k\!\left(3/2\right).
\label{eq:BdGd}
\end{equation}
We replace $u_k\!\left(-1\right)$, $u_k\!\left(0\right)$,
$u_k\!\left(1\right)$, $v_k\!\left(3/2\right)$, $u_k\!\left(2\right)$
in (\ref{eq:BdGe}-\ref{eq:BdGd}) with $\Psi_{1L}\!\left(-1\right)$,
$\Psi_{1L}\!\left(0\right)$, $\Psi_{1R}\!\left(1\right)$,
$\Psi_{2R}\!\left(3/2\right)$, $\Psi_{1R}\!\left(2\right)$,
respectively, obtaining a linear system for the three unknown quantities
$v_k\!\left(1/2\right)$, $R_k$, and $T_k$:
\begin{equation}
\cases{
-\left(\tilde{\Delta}_L+\tilde{V}_{dfL}\right)R_k+\left(\omega-
\varepsilon_{f1/2}\right)v_k\!\left(1/2\right)+
\left(\tilde{\Delta}_R-\tilde{V}_{dfR}\right)u_k{\rm e}^{{\rm i}ka}T_k=
-\left(\tilde{\Delta}_L+\tilde{V}_{dfL}\right)& \cr
-tR_k+\left(\tilde{\Delta}_R^*-\tilde{V}_{dfR}^*\right)
v_k\!\left(1/2\right)+\left(\omega u_k{\rm e}^{{\rm i}ka}+t u_k
{\rm e}^{2{\rm i}ka}+\left(\Delta^*+V_{df}^*\right)v_k
{\rm e}^{(3/2){\rm i}ka}\right)T_k=-t& \cr
\left(-\omega+\varepsilon_{d0}-t^{\prime}{\rm e}^{{\rm i}qa}\right)R_k
+\left(\tilde{\Delta}^*_L+\tilde{V}^*_{dfL}\right)
v_k\!\left(1/2\right)+tu_k{\rm e}^{{\rm i}ka}T_k=-\omega+\varepsilon_{d0}
-t^{\prime}{\rm e}^{-{\rm i}qa}.& \cr
} 
\label{eq:system}
\end{equation}
To solve system (\ref{eq:system}) we fix $\omega$, 
then we invert Eqs.~(\ref{eq:omegap}) and (\ref{eq:energyl})
to obtain $k$ and $q$, and hence $u_k$ and $v_k$: 
once we input as parameters 
$\tilde{\Delta}_L$, $\tilde{\Delta}_R$, $\tilde{V}_{dfL}$,
$\tilde{V}_{dfR}$, $\varepsilon_{d0}$, and $\varepsilon_{f1/2}$,
the whole coefficient matrix of system
(\ref{eq:system}) is known and $T_k$ and $R_k$ can be eventually
obtained.
\end{widetext}
In Eq.~(\ref{eq:incoming}) we chose a normalization such that the
flux of the incident wave $J_{keN}^{\rm inc}\!\left(j\right)$ is 
\begin{equation}
J_{keN}^{\rm inc}\!\left(j\right)=
\frac{2t^{\prime}}{\hbar}\sin{\left(qa\right)}=
\frac{1}{a\hbar}\frac{\partial\,\omega^{\prime}\!\left(q\right)}{\partial q}
\quad \forall j<0,
\end{equation}
and that of the reflected wave  $J_{keN}^{\rm refl}\!\left(j\right)$
is
\begin{equation}
J_{keN}^{\rm refl}\!\left(j\right)=
-\left|R_k\right|^2\frac{2t^{\prime}}{\hbar}\sin{\left(qa\right)}
\quad \forall j<0,
\end{equation}
hence, by definition, the reflection coefficient 
\protect{${\cal{R}}\!\left(\omega^{\prime}
\!\left(q\right)\right)={\cal{R}}\!\left(\omega
\!\left(k\right)\right)$} is given by
\begin{equation}
{\cal{R}}\!\left(\omega\right)=\frac{\left|J_{keN}^{\rm refl}\right|}
{\left|J_{keN}^{\rm inc}\right|}=\left|R_k\right|^2.
\label{eq:rdef}
\end{equation}
The transmission coefficient ${\cal{T}}\!\left(\omega\right)$
can be calculated by the definition
\begin{equation}
{\cal{T}}\!\left(\omega\right)=\frac{\left|J_{keN}^{\rm trans}\right|}
{\left|J_{keN}^{\rm inc}\right|},
\label{eq:tdef}
\end{equation}
with
\begin{eqnarray}
&& J_{keN}^{\rm trans}\!\left(j\right)=
\frac{2t}{\hbar}
\left|T_k\right|^2 u_k^2\sin{\left(ka\right)}
+\frac{1}{\hbar}\left|T_k\right|^2
u_k\left|v_k\right| \nonumber\\
&&\times\quad \Big[\left(\Delta+V_{df}\right)
\sin{\left(ka/2+\varphi\right)}\nonumber\\
&&\qquad+\left(\Delta-V_{df}\right)
\sin{\left(ka/2-\varphi\right)}\Big]\quad 
\forall j>0,
\end{eqnarray}
or, more simply, by probability conservation
\begin{equation}
{\cal{T}}\!\left(\omega\right)=1-{\cal{R}}\!\left(\omega\right).
\label{eq:conserv}
\end{equation}
${\cal{T}}\!\left(\omega\right)$ 
is the key quantity we need to compute the thermopower.

Because of the two-fold degeneracy of energy $\omega$, we have to
consider also the quasi-particle associated with $-k$ (and $-q$), with
$k>0$. It is clear how to rewrite 
Eq.~(\ref{eq:incoming}-\ref{eq:transmitted}),
because now the electron, coming from the right, is partly 
reflected into the right-hand side of the junction and 
partly transmitted into the left-hand side, i.e.
\begin{equation}
{\Psi_{1L}\!\left(j\right) \choose \Psi_{2L}\!\left(j+1/2\right)}
=T_{-k}{1 \choose 0}{\rm e}^{-{\rm i}qaj},
\end{equation}
\begin{eqnarray}
&& {\Psi_{1R}\!\left(j\right) \choose \Psi_{2R}\!\left(j+1/2\right)}
={u_{-k} \choose v_{-k}{\rm e}^{-{\rm i}ka/2}}{\rm e}^{-{\rm i}kaj} 
\nonumber\\ &&\qquad -\quad 
R_{-k}{u_{-k} \choose v_{-k}{\rm e}^{{\rm i}ka/2}}{\rm e}^{{\rm i}kaj}.
\end{eqnarray}
After that, we can proceed in the same way as above. By
time-reversal symmetry, one has
\begin{equation}
{\cal{T}}\!\left(\omega\!\left(k\right)\right)=
{\cal{T}}\!\left(\omega\!\left(-k\right)\right).
\end{equation}
An analogous ER procedure is used to compute the transmission and
reflection coefficients $\bar{{\cal{T}}}\!\left(\bar{\omega}\right)$
and $\bar{{\cal{R}}}\!\left(\bar{\omega}\right)$ for holes. In this last case,
Eqs.~(\ref{eq:incoming}-\ref{eq:transmitted}) turn into (with $q,k>0$)
\begin{equation}
{\bar{\Psi}_{1L}\!\left(j\right) \choose \bar{\Psi}_{2L}\!\left(j+1/2\right)}
={1 \choose 0}{\rm e}^{-{\rm i}qaj}-
\bar{R}_{-k}{1 \choose 0}{\rm e}^{{\rm i}qaj},
\end{equation}
\begin{equation}
{\bar{\Psi}_{1R}\!\left(j\right) \choose \bar{\Psi}_{2R}\!\left(j+1/2\right)}
=\bar{T}_{-k}{\bar{u}_{-k} \choose \bar{v}_{-k}
{\rm e}^{-{\rm i}ka/2}}{\rm e}^{-{\rm i}kaj}.
\end{equation}

The above derivation of the transmission coefficient still holds 
for case (iii) as long as we rename and set the parameters as discussed
in section \ref{model}. Some caution is needed
in taking the correct sign of the wavevector $k$
for certain values of parameters outside the range we actually
examined.
Namely, if \protect{$2t^2(\Delta^2+V_{df}^2)<(\Delta^2
-V_{df}^2)^2$} and \protect{$t^2<\Delta^2-V_{df}^2$} 
the electron wavefunction (\ref{eq:provap}) travels from right to left if
$k>0$, like the hole wavefunction (\ref{eq:provam}) when 
\protect{$2t^2(\Delta^2+V_{df}^2)<(\Delta^2
-V_{df}^2)^2$} and \protect{$t^2<V_{df}^2-\Delta^2$}.

\subsection{Transport coefficients and $ZT$}\label{trasporto}

If we apply an electric field $E$ or a temperature gradient
$\nabla T$ across the junction, we produce electric and thermal currents 
$J_E$ and $J_T$, respectively. In a stationary state the
relation between fluxes and driving forces is linear,
if fields are small enough:\cite{ziman}
\begin{equation}
\cases{J_E=L_{EE}E+L_{ET}\nabla T & \cr
J_T=L_{TE}E+L_{TT}\nabla T. & \cr}
\label{eq:ziman1}
\end{equation}
The transport coefficients $L$ are not all independent, because of
the Onsager relation
\begin{equation}
L_{ET}=-\frac{L_{TE}}{T}.
\label{eq:onsager}
\end{equation}
The coefficients are related  to the
electrical and thermal conductivities $\sigma$ and $\kappa$,\cite{kappa}
respectively,
and to the thermopower (Seebeck coefficient) $Q$ of the junction:
\begin{eqnarray}
\sigma=L_{EE}, && \kappa=-\left(L_{TT}-\frac{L_{TE}L_{ET}}{L_{EE}}\right),
\nonumber\\ Q&=&-\frac{L_{ET}}{L_{EE}}.
\label{eq:rel1}
\end{eqnarray}
A high value of $Q$ alone does not necessarily imply that the junction is a
good thermoelectric cooler, that is why 
engineers introduce the figure of merit $Z$:
\begin{equation}
Z=\frac{Q^2\sigma}{\kappa}.
\label{eq:rel2}
\end{equation}
Indeed, if we pump heat from the cooler side to the hotter one, we need high
pumping efficiency (high $Q$), low production of heat through Joule heating
(high $\sigma$), and low backwards conduction of heat 
(low $\kappa$).\cite{optimumgap}
$Z$ has units of inverse temperature, thus in general what is quoted is
the dimensionless quantity $ZT$. We shall also use another natural
definition of figure of merit, $ZT/\left(1+ZT\right)$, which is
the dimensionless figure, defined below in Eq.~(\ref{eq:alternative}), 
more directly related to the transmission coefficient momenta we compute.

Since the computation of the electrostatic and thermal field
across the junction is complicated and not essential to the
junction thermopower, 
we simply assume that the voltage $V$ and the temperature $T$ are
constant on both sides of the junction and have a sharp step
at the interface. Thus, instead of Eq.~(\ref{eq:ziman1}), we write 
\begin{equation}
\cases{J_E=K_{EE}\left(-\delta V\right)+K_{ET}\,\delta T & \cr
J_T=K_{TE}\left(-\delta V\right)+K_{TT}\,\delta T, & \cr}
\label{eq:ziman2}
\end{equation}
where we have replaced the gradients $E=-\nabla V$ and $\nabla T$
with $\left(-\delta V\right)$ and $\delta T$, respectively,
where \protect{$\delta V=V\left({\rm right}\right)-V\left({\rm left}\right)$}
and \protect{$\delta T=T-T_n$}, being $T$ the temperature associated to 
the right-hand side, and $T_n$ to the normal metal on the left-hand side.
Since we have assumed \protect{$V\left({\rm right}\right)=0$},
we have \protect{$-\delta V=V\left({\rm left}\right)=V$}.
The coefficient $K_{EE}$ represents now the conductance $G$ in place of the 
conductivity $\sigma$ and \protect{$-\left(K_{TT}-
K_{TE}K_{ET}/K_{EE}\right)$} the thermal conductance 
$G_T$\cite{condterm} in place of the thermal conductivity $\kappa$, but all the
other relations (\ref{eq:onsager}-\ref{eq:rel2}) still hold: 
\begin{equation}
K_{ET}=-\frac{K_{TE}}{T},
\label{eq:ons2}
\end{equation}
\begin{eqnarray}
G=K_{EE}, && G_T=-\left(K_{TT}-\frac{K_{TE}K_{ET}}{K_{EE}}\right),
\nonumber\\
Q &=& -\frac{K_{ET}}{K_{EE}},
\label{eq:rela1}
\end{eqnarray}
\begin{equation}
Z=\frac{Q^2G}{G_T}.
\label{eq:rela2}
\end{equation}
Note that the Onsager relation (\ref{eq:ons2})
is still valid despite our simplifing
assumptions about the $V$- and $T$-gradients.
From Eq.~(\ref{eq:ziman2}) the  
formulae for the $K$ transport coefficients follow:  
\begin{eqnarray}
K_{EE}&=&\left(\frac{\partial J_E}{\partial \left(-\delta V\right)}
\right)_{\delta T, \delta V=0},\nonumber\\
K_{ET}&=&\left(\frac{\partial J_E}{\partial \left(\delta T\right)}
\right)_{\delta T, \delta V=0},\nonumber\\
K_{TE}&=&\left(\frac{\partial J_T}{\partial \left(-\delta V\right)}
\right)_{\delta T, \delta V=0},\nonumber\\
K_{TT}&=&\left(\frac{\partial J_T}{\partial \left(\delta T\right)}
\right)_{\delta T, \delta V=0}.
\label{eq:kdef}
\end{eqnarray}

Here we closely follow the approach of Blonder {\em et al.}\cite{BTK}
We assume that the two sides of the junction are in contact 
with perfect electron reservoirs,
at different temperatures and chemical potentials, and that
quasi-particle wavefunctions keep their phase
coherence across the whole system except at $z = \pm \infty$, where they 
completely loose their phase, thermalize and relax in energy due to  
inelastic scattering processes of the Fermi sea. Namely, we regard
the contacts as perfect emitters or adsorbers.
The electric current density, $J_E$, will be given by the sum of 
all contributions of the quasi-particle
excitations to the current, each one weighted
by the correct Fermi distribution function, depending if
quasi-particles originate from the left or right reservoir.
Because $J_E$ is stationary and 
conserved, we can calculate it in every point of the space, 
so we choose a lattice site $j<0$ in the bulk on the left-hand side. 
In particular, the contribution to the density current $J_{E}^{L\rightarrow R}$
produced by electrons going from left to right is [see Eq.~(\ref{eq:counting})] 
\begin{equation}
J_{E}^{L\rightarrow R}=2\frac{1}{N_s}\sum_{k>0}J_{ke}^{\rm inc}\!\left(
j\right)f\!\left(\omega\!\left(k\right)-eV\right)\quad j<0,
\label{eq:lr}
\end{equation}
where the Fermi function refers to the temperature $T_n$ of the metal on 
the left-hand side, and the electrochemical potential differs from that 
on the right-hand side by $eV$. The prefactor $2$ accounts for the spin 
degeneracy. Note that Eq.~(\ref{eq:lr}) gives the only electron flux
running from left to right on the bulk metal side.
On the other hand, the contribution to the electron 
current $J_{E}^{L \leftarrow R}$ from right to left,
always on the bulk metal side, is given by
two distinct terms, one for electrons reflected at
the interface and coming from the left side, hence in equilibrium with
the left reservoir (with temperature $T_n$ and Fermi function referred
to $eV$), and one for electrons transmitted through the
interface and coming from the right reservoir at temperature $T$ and
at ground potential:
\begin{eqnarray}
J_{E}^{L\leftarrow R} &=& 2\frac{1}{N_s}\sum_{k>0}J_{ke}^{\rm refl}\!\left(
j\right)f\!\left(\omega\!\left(k\right)-eV\right) \nonumber\\
&+& 2\frac{1}{N_s}\sum_{k<0}J_{ke}^{\rm trans}\!\left(
j\right)f\!\left(\omega\!\left(k\right)\right)
\quad j<0.
\label{eq:rl}
\end{eqnarray}
Note that the second sum in Eq.~(\ref{eq:rl}) runs over negative
$k$ values and the related electron wavefunctions extend all over
the junction, being therefore well defined at $j<0$.
Now we add (\ref{eq:lr}) and (\ref{eq:rl}), using
(\ref{eq:rdef}) and (\ref{eq:tdef}),
\begin{eqnarray}
&& J_{E}^{L\rightarrow R}+J_{E}^{L\leftarrow R}=
2\frac{1}{N_s}\sum_{k>0}J_{ke}^{\rm inc}\!\left(
j<0\right) \nonumber\\
&&\times\quad \left[1-{\cal{R}}\!\left(\omega\!\left(k\right)\right)\right]
f\!\left(\omega\!\left(k\right)-eV\right)\nonumber\\
&&+\quad 2\frac{1}{N_s}\sum_{k<0}J_{ke}^{\rm inc}\!\left(j>0
\right){\cal{T}}\!\left(\omega\!\left(k\right)\right)
f\!\left(\omega\!\left(k\right)\right).
\end{eqnarray}
Here the notation \protect{$J_{ke}^{\rm inc}\!\left(j>0
\right)$} refers to electrons, incident
on the interface, coming from $z=+\infty$.
From Eqs.~(\ref{eq:semiclassic}) and (\ref{eq:semiclassic2})
\begin{eqnarray}
&&J_{E}^{L\rightarrow R}+J_{E}^{L\leftarrow R}=
2\frac{1}{N_s}\sum_{q>0}
\frac{e}{a\hbar}
\frac{\partial\, \omega^{\prime}\!\left(q\right)}{\partial q} \nonumber\\
&&\times\quad
\left[1-{\cal{R}}\!\left(\omega^{\prime}\!\left(q\right)\right)\right]
f\!\left(\omega^{\prime}\!\left(q\right)-eV\right)\nonumber\\
&&+\quad 2\frac{1}{N_s}\sum_{k<0}
\frac{e}{a\hbar}
\frac{\partial\, \omega\!\left(k\right)}{\partial k}
{\cal{T}}\!\left(\omega\!\left(k\right)\right)
f\!\left(\omega\!\left(k\right)\right),
\end{eqnarray}
and going to the continuum limit \protect{$\sum_k\rightarrow 
\left(N_sa\right)/\!\left(2\pi\right)\int{\rm d\,}k$} we obtain
\begin{eqnarray}
&& J_{E}^{L\rightarrow R}+J_{E}^{L\leftarrow R}=
\frac{2e}{2\pi\hbar}\int_0^{\pi/a}\!\!\!{\rm d\,}q \nonumber\\
&&\times\quad \frac{\partial\, \omega^{\prime}\!\left(q\right)}{\partial q}
{\cal{T}}\!\left(\omega^{\prime}\!\left(q\right)\right)
f\!\left(\omega^{\prime}\!\left(q\right)-eV\right)\nonumber\\
&&+\quad \frac{2e}{2\pi\hbar}\int_{-\pi/a}^0\!\!\!{\rm d\,}k
\frac{\partial\, \omega\!\left(k\right)}{\partial k}
{\cal{T}}\!\left(\omega\!\left(k\right)\right)
f\!\left(\omega\!\left(k\right)\right),
\end{eqnarray}
or
\begin{eqnarray}
&& J_{E}^{L\rightarrow R}+J_{E}^{L\leftarrow R} =
\frac{2e}{h}\int_0^{\infty}\!\!\!{\rm d\,}\omega \nonumber\\
&&\times\quad {\cal{T}}\!\left(\omega\right)\left[f\!\left(\omega
-eV\right)-f\!\left(\omega\right)\right],
\label{eq:adde}
\end{eqnarray}
with the convention that \protect{${\cal{T}}\!\left(\omega\right)=0$}
if $\omega$ is not in the range of excitation energies allowed,
and \protect{$h=2\pi\hbar$}.
Similarly, the hole contribution to $J_E$ is given by
\begin{equation}
-\frac{2e}{h}\int_0^{\infty}\!\!\!{\rm d\,}\bar{\omega}
\bar{{\cal{T}}}\!\left(\bar{\omega}\right)\left[f\!\left(\bar{\omega}
+eV\right)-f\!\left(\bar{\omega}\right)\right].
\label{eq:addh}
\end{equation}
Now we add (\ref{eq:adde}) and (\ref{eq:addh}),
passing to SCR and using the equivalence
\protect{$f\!\left(-\epsilon\right)=
1-f\!\left(\epsilon\right)$}, to obtain the current $J_E$:
\begin{eqnarray}
J_E &=& \frac{2e}{h}\int_{-\infty}^{\infty}\!\!\!{\rm d\,}
\epsilon\left[{\cal{T}}\!\left(\epsilon\right)+
\bar{{\cal{T}}}\!\left(-\epsilon\right)\right] \nonumber\\
&&\times\quad
\left[f\!\left(\epsilon-eV\right)-f\!\left(\epsilon\right)\right].
\label{eq:Je}
\end{eqnarray}
The electron (hole) heat current density \protect{$J_{keT}\!\left(j\right)$}
[\protect{$J_{-khT}\!\left(j\right)$}] is defined by:
\begin{eqnarray}
J_{keT}\!\left(j\right) &=& \omega\!\left(k\right)J_{keN}\!\left(j\right),
\nonumber\\
J_{-khT}\!\left(j\right) &=& \bar{\omega}
\!\left(-k\right)J_{-khN}\!\left(j\right)
\end{eqnarray}
(note that in ER energies $\omega$ are referred to $\mu$).
Reasoning in the same way as for $J_E$, we obtain the total heat current
$J_T$:
\begin{eqnarray}
J_T &=& \frac{2}{h}\int_{-\infty}^{\infty}\!\!\!{\rm d\,}
\epsilon\left[{\cal{T}}\!\left(\epsilon\right)+
\bar{{\cal{T}}}\!\left(-\epsilon\right)\right]
\nonumber\\ &&\times\quad \epsilon
\left[f\!\left(\epsilon-eV\right)-f\!\left(\epsilon\right)\right].
\end{eqnarray}
Now it is easy to compute $K$-transport coefficients from definitions
(\ref{eq:kdef}). We only summarize the results:
\begin{eqnarray}
K_{EE}=\frac{2e^2}{h}L_0,&\qquad\qquad&
K_{ET}=-\frac{1}{T}\,\frac{2e}{h}L_1,\nonumber\\
K_{TE}=\frac{2e}{h}L_1,&\qquad\qquad&
K_{TT}=-\frac{1}{T}\,\frac{2}{h}L_2,
\label{eq:ksolution}
\end{eqnarray}
with
\begin{equation}
L_n=
\int_{-\infty}^{\infty}\!\!\!{\rm d\,}
\epsilon\left[{\cal{T}}\!\left(\epsilon\right)+
\bar{{\cal{T}}}\!\left(-\epsilon\right)\right]\epsilon^n
\left(-\frac{\partial\,f\!\left(\epsilon\right)}
{\partial\,\epsilon}\right).
\label{eq:ldef}
\end{equation}
From Eq.~(\ref{eq:ksolution}) one can immediately check that
the Onsager relation (\ref{eq:ons2}) is fulfilled. Besides, note
the striking similarity of Eq.~(\ref{eq:ksolution}-\ref{eq:ldef})
with the formalism of the semiclassical theory
of transport (Boltzmann Equation):\cite{ziman} in our case the role
of the \protect{$k$-dependent} relaxation time $\tau\!\left(k\right)$ is
played by the transmission coefficient ${\cal{T}}$, 
the interface being the
scattering mechanism.   
From the transmission coefficient momenta (\ref{eq:ldef}) we compute the 
alternative figure of merit
\begin{equation}
\frac{L_1^2}{L_0L_2}=\frac{ZT}{ZT+1}.
\label{eq:alternative}
\end{equation}

\section{Results and discussion}\label{risultati}

We present the results for the thermopower $Q$ and the figure of merit
$ZT$ for the different types of junction under study.
Bulk parameters are chosen to describe a large
class of materials and are the same as in 
Fig.~\ref{fig2}, with the metal band much broader than
the semiconductor one. In particular, for FE and NB 
on the right-hand side of the junction the 
indirect gap $E_{\rm gap}$ at $T=0$ is 0.40$t$, i.e.~one tenth of the 
``bare'' $d$-bandwidth 4$t$ [\protect{$E_{\rm gap}=-t+
\sqrt{t^2+4\Delta^2\!\left(0\right)}$} and 
\protect{$-t+\sqrt{t^2+4V_{df}^2}$}, respectively].
In case (iii), the gap is direct, 
\protect{$E_{\rm gap}=$ 0.40$t$}, where 4$t$ is roughly
the total bandwidth (electron plus hole band) of SC
[\protect{$E_{\rm gap}=2\sqrt{t^{\prime\prime 2}+4V_{\text{e}}^2}$}].
With this choice of parameters, the three semiconductors have the same gap
at $T=0$ and approximately the same bandwidth.

\subsection{Clean interface}\label{pulita}

First we study the junction with a clean interface. With this 
terminology we mean that there are no $d$- or $f$-impurity layers 
(\protect{$\varepsilon_{d0}=\varepsilon_{f1/2}=0$}) at $z=0$ or
$z=a/2$, respectively. For case (i) we assume that
the change in the order parameter 
$\Delta\!\left(T\right)$ at $z=0$ is
abrupt, from zero to the bulk value [\protect{$\tilde{\Delta}_L=
\tilde{\Delta}_R=\Delta\!\left(T\right)$}]. 
Similarly, we take the hybridization or hopping
parameters at the interface equal to the bulk values in both cases (ii)
[\protect{$\tilde{V}_{dfL}=\tilde{V}_{dfR}=\tilde{V}_{df}$}]
and (iii) 
[\protect{$\tilde{V}_{eL}=\tilde{V}_{eR}=V_{\text{e}}$}, 
\protect{$\tilde{t}_{L}=\tilde{t}_{R}=t$}].

\begin{figure}
\begin{picture}(300,420)
\put(-36,-50)
{\epsfig{file=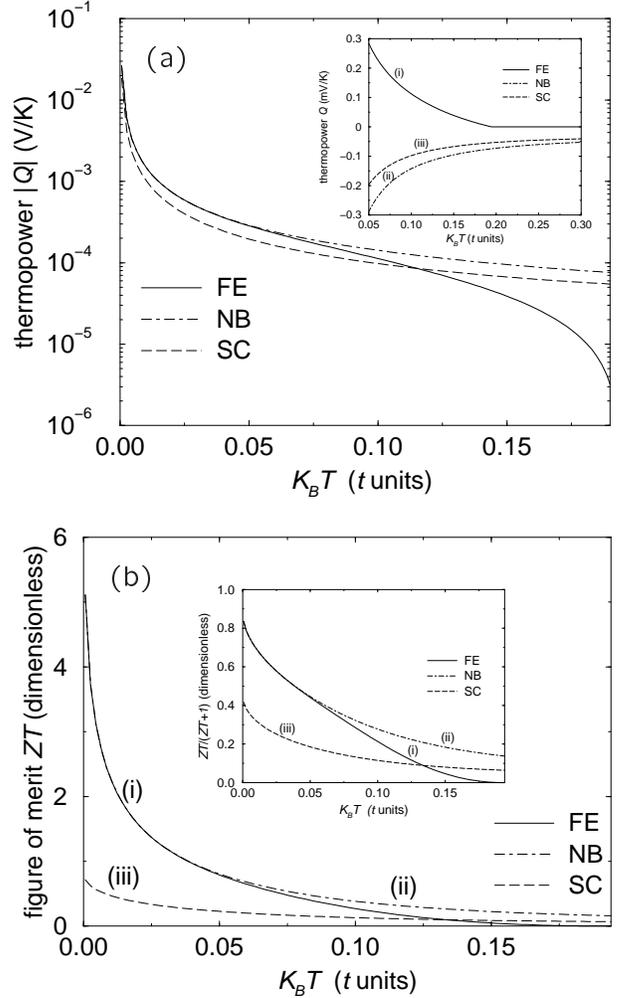,width=4.4in,,angle=0}}
\end{picture}
\caption{
(a) Absolute value of the thermopower $\left|Q\right|$ vs $T$,
for the three cases under study. $\left|Q\right|$ is in logarithmic scale.
Note that $\left|Q\right|$ of FE drops to zero, as
\protect{$T\rightarrow T_c$} (\protect{$T_c=$ 0.195$t$}, with parameters of
Fig.~\ref{fig2}). Inset: Magnification of the same plot, in linear scale,
in the neighborhood of $T_c$ for FE.
If $T>T_c$ then $Q=0$ for FE, because above the critical temperature
the gap vanishes and the bulk FE
turns into a metal [\protect{${\cal{T}}\!\left(\epsilon\right)
=\bar{{\cal{T}}}\!\left(-\epsilon\right)$}].
(b) Plot of the corresponding figure of merit $ZT$ vs $T$.
Like $Q$ in Fig.~\ref{fig3}(a), $ZT$ of FE goes to zero as
\protect{$T\rightarrow T_c$}. Inset: same plot for the alternative figure
of merit \protect{$ZT/\left(ZT+1\right)$}, whose upper bound
(ideal thermoelectric) is 1.
}
\label{fig3}
\end{figure}
In Fig.~\ref{fig3}(a) we plot the absolute value of
thermopower $\left|Q\right|$ vs temperature.
In all three cases $\left|Q\right|$ goes to infinity 
as \protect{$T\rightarrow 0$},
as expected for both indirect-gap narrow-band semiconductors\cite{castro} 
and direct-gap ordinary semiconductors.\cite{ziman}
As $T$ approaches the critical temperature $T_c$ of the ferroelectric
(\protect{$k_{\text{B}}T_c=$0.195$t$} for our choice of parameters), 
the thermopower $\left|Q\right|$ of FE
goes to zero, while $\left|Q\right|$ for NB and SC decreases 
in a exponential-like manner as the temperature rises. 
At $T=T_C$ the FE gap vanishes and the semiconductor
turns into a metal with symmetric bands with respect to $\mu$,
i.e.~$Q=0$. In the low temperature
region, instead, for \protect{$k_{\text{B}}T\leq$ 0.05$t$} 
(\protect{$T\leq$ 0.25$T_c$}),
i.e., in the region where $\Delta\!\left(T\right)$ is almost constant,
far from the second-order ferroelectric/metal transition,
the absolute value of $Q$ for FE and NB is 
nearly identical. For SC  
$\left|Q\right|$ is about \protect{60\%-70\%} 
of the NB value in the whole range
of temperature. To compare our results with reported bulk data, we can 
assume a typical value of \protect{4$t=$ 1 eV} for FE or NB, i.e.~a gap
around \protect{0.1 eV} at $T=0$, with room temperature 
\protect{$\sim 0.1 t$.} With these
numbers, we have \protect{$\left|Q\right|=$ 0.11 -- 0.14 mV/K} 
at room temperature, values comparable with those of some rare-earth metals
with very high thermopower.
If instead we assume for the broad band semiconductor SC a typical gap of
\protect{0.5 eV,} then $\left|Q\right|$ is around 
\protect{0.5 mV/K} at room
temperature, consistently with 
characteristic bulk data. 
The inset of Fig.~\ref{fig3}(a) is a magnification of the
plot (in linear scale) in the neighborhood of $T_c$. The discontinuity
of the derivative of $Q$ for the FE curve 
is a signature of the second-order
ferroelectric transition: if $T>T_c$ then $Q=0$. Similar kinks are
found also for the other transport coefficients $G$ and $G_T$.
One sees that NB and SC have electron transport character ($Q<0$)
while FE has hole character ($Q>0$).

\begin{figure}
\centerline{\epsfig{file=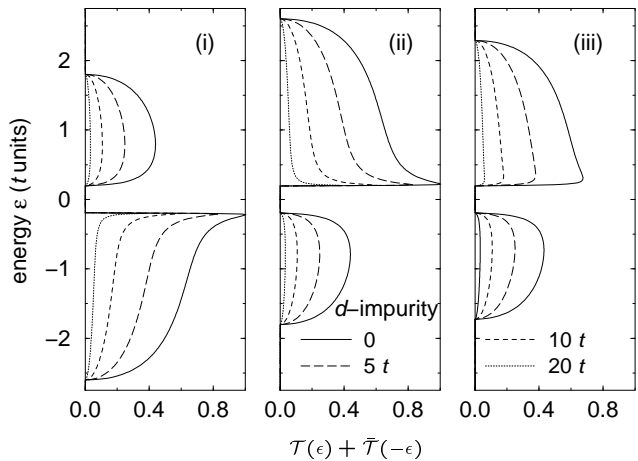,width=2.6in,,angle=90}}
\caption{
Total (electron plus hole) transmission coefficient
\protect{${\cal{T}}\!\left(\epsilon\right)
+\bar{\cal{T}}\!\left(-\epsilon\right)$}
in ``semiconductor representation'' (SCR) as a function of the energy
$\epsilon$ (vertical axis). From left to right
panels correspond to FE, NB, and SC,
respectively. Curves for different values of
the $d$-impurity energy $\varepsilon_{d0}$ are plotted. Note that
\protect{${\cal{T}}\!\left(\epsilon\right)$} of FE is mapped into
\protect{$\bar{\cal{T}}\!\left(-\epsilon\right)$} of NB under the
transformation \protect{$\epsilon\rightarrow -\epsilon$}, as a consequence
of equations (\ref{eq:canonical}). The $d$-impurity
energy for four curves is indicated in the diagram above.
Other parameters used are the
same as in Fig.~\ref{fig2} and $T=0$.
}
\label{fig4}
\end{figure}
To understand the above features, in Fig.~\ref{fig4} we plot the
total (electron plus hole) transmission coefficient 
\protect{${\cal{T}}\!\left(\epsilon\right)+
\bar{\cal{T}}\!\left(-\epsilon\right)$} vs energy in SCR for the
three cases under study (at $T=0$). 
Solid lines refer to the clean interface case. Since the sign 
and magnitude of $Q$ is established by Eq.~(\ref{eq:ldef}) for $L_1$, i.e.~by
the competition between the different weights of electron and hole
transmission functions $\cal{T}$ and $\bar{\cal{T}}$, respectively,
it is clear that the stronger the electron/hole asymmetry of the
transmission coefficient, the higher the thermopower.  
Panels (i) and (ii) of Fig.~\ref{fig4} present the remarkable electron/hole
asymmetry close to the gap, hence
$Q$ for FE has a strong hole character ($Q>0$) while
NB has a dominant electron character ($Q<0$). The mapping of
${\cal{T}}\!\left(\epsilon\right)$ of FE into 
$\bar{\cal{T}}\!\left(-\epsilon\right)$ of NB under the transformation
\protect{$\epsilon\rightarrow -\epsilon$} 
---a consequence of the transformation
(\ref{eq:canonical})--- explains why $\left|Q\right|$ is
the same for FE and NB at low $T$. The situation is different for
SC [see panel (iii)], because electron and hole 
transmission coefficients are more symmetric, and thus
$\left|Q\right|$ has a lower value. In the limit of electron/hole symmetry 
[\protect{${\cal{T}}\!\left(\epsilon\right)
=\bar{\cal{T}}\!\left(-\epsilon\right)$}, as it is the case for FE when 
\protect{$T\rightarrow T_c$}], the thermopower is zero.

The other transport parameters, apart from $Q$, 
are the electrical conductance $G$ and the
thermal conductance $G_T$. We find that, for all the three cases under
study, there exists an activation temperature around
0.03$t$, below which $G$ and $G_T$, as functions of $T$, 
rapidly (exponentially) drop to zero,
because the number of thermally excited carriers becomes too small.
This behavior, characteristic of a material with gap, shows 
that at low temperature the main contribution to the thermal
conductance $G_T$ is given by the lattice, which is not
included, thus dramatically decreasing
the actual value of $ZT$. This contribution,
and hence the minimum working temperature of the junction, depends
on the thermal impedance mismatch of the interface.

Once all transport coefficients are known,
the most relevant quantitity to be computed for practical applications
is the figure
of merit $ZT$, plotted in Fig.~\ref{fig3}(b) as a function of the temperature.
In all three cases $ZT$ is a monotonic decreasing function of $T$. However,
$ZT$ is much bigger for FE and NB than
for SC (but if \protect{$T\rightarrow T_c$} then \protect{$ZT
\rightarrow 0$} for FE). With the numerical parameters we have employed above,
at room temperature $ZT$ is \protect{$\sim\,$0.3} for FE 
and \protect{$\sim\,$0.4} for NB,
but already at \protect{$T=$100 K} 
$ZT$ is \protect{$\sim\,$1.1} for FE and NB and
only \protect{$\sim\,$0.5} for SC.
In the inset of Fig.~\ref{fig3}(b), we redraw the same plot in term of the
alternative figure of merit \protect{$ZT/\left(ZT+1\right)$}: this is
the quantity we will examine in the following. While $ZT$ has no
theoretical upper bound, the maximum of $ZT/(ZT+1)$ is 1,
corresponding to $ZT=\infty$.

It is likely that the one-dimensional model artificially enhances $ZT$. 
Besides, we do not know how the order parameter $\Delta\!\left(T\right)$
actually varies at the interface, and all the effects of charge polarization
are neglected: this does not necessarily imply a reduction of $ZT$.
The only scattering mechanism we have considered is the interface: 
in particular, the contribution
of phonons to the thermal conductance is neglected.

\subsection{$d$-impurity layer}\label{sporcad}

Now we study the effect of an overlayer of $d$-impurity atoms at the 
interface. This structure could be built by epitaxial growth techniques: 
we find that this configuration does not
improve considerably the figure of merit $ZT$.

In our one dimensional virtual crystal model, we describe 
for simplicity the overlayer 
by putting one atom at $z=0$ with site energy $\varepsilon_{d0}$. 
Since in the calculation we leave unchanged
the hopping coefficient $t$ and all the other parameters 
with respect to the clean-interface case, 
we also require, from a physical point of view, that the atomic orbital
of the impurity is still of $d$-type. The
essential point here is that the atom at $z=0$ substantially participates
in the electronic motion, contrary to a $f$-site. In case (iii)
this distinction between $f$- and $d$-sites is no longer relevant: 
actually we look at SC only for comparison.
One could also think of growing an overlayer
at $z=0$ only partially filled with impurities: 
in that case the values of transport coefficients should be an 
interpolation between the two limiting values corresponding to the cases of
0\% (clean surface) and 100\% impurity concentration within the layer.  

To understand the role of the $d$-impurity layer in the transport, 
consider the transmission in Fig.~\ref{fig4}. Here, as well as the 
total transmission coefficient vs energy
for the clean interface (solid lines), we have also plotted curves
corresponding to increasing values of
the impurity level $\varepsilon_{d0}$. 
As the impurity level energy $\varepsilon_{d0}$ increases, 
the transmission is uniformly depressed over the whole range of
energies. In case (i) and (ii), results depend only on the absolute
value of $\varepsilon_{d0}$, while (iii) is
more complex, due to the presence of a second nearest neighbor hopping
coefficient.

While the depression effect of the impurity 
seems important for both NB and FE semiconductors, 
some caution is needed in accepting these results. 
First, one cannot make $\left|\varepsilon_{d0}\right|$ arbitrarily large,
because its value is physically limited and cannot differ too much from
typical bulk energies, otherwise the impurity would be screened. 
Second, in our computation we have taken 
all parameters but $\varepsilon_{d0}$
unchanged with respect to the clean-interface case, and it is clear that
this approximation becomes worse as long as $\left|\varepsilon_{d0}\right|$
increases. For example, $\tilde{\Delta}_l$ or $\tilde{V}_{dfL}$
would surely change as $\varepsilon_{d0}$ varies.

The conclusion is that the enhancement of the figure of merit
\protect{$ZT/\left(ZT+1\right)$} 
for reasonable values of $\varepsilon_{d0}$ is quite limited.
In general, as long as we increase
the magnitude of $\varepsilon_{d0}$, we 
slightly enhance $ZT$
uniformly over the whole range of temperatures. 
It is remarkable that FE and NB values of $ZT$ are always much higher 
than SC values.
For example, at room temperature, 
with the usual choice of numerical parameters
and $\varepsilon_{d0}=5t$, 
\protect{$ZT\sim\,$0.4} for FE,
\protect{$\sim\,$0.6} for NB, and \protect{$\sim\,$0.2} for SC, 
but already at \protect{100 K} \protect{$ZT\sim\,$1.8} 
for FE and NB, while it is only
\protect{$\sim\,$0.4} for SC. 

\subsection{$f$-impurity layer}\label{sporcaf}

In order to dramatically improve the figure of merit, we propose the 
insertion of a rare-earth overlayer at $z=a/2$, with $f$-level energy 
$\varepsilon_{f1/2}$. Contrary to the $d$-impurity layer discussed in 
the previous section, here it is essential that the localized impurity 
level is of $f$-type, i.e.~hardly sharing the electronic conduction. 
This is not the case for SC, where tunneling from this ``$f$-site''
to adjacent neighbors is allowed: in fact, for SC,
the situation is basically similar to the previous 
$d$-impurity case. In case (iii), hopping to adjacent neighbors is allowed
both from the ``$f$-site'' and from the ``$d$''-site.
In fact, SC results for the figure of merit $ZT$ are not dissimilar from
the values obtained in the $d$-impurity case.
\begin{figure}
\centerline{\epsfig{file=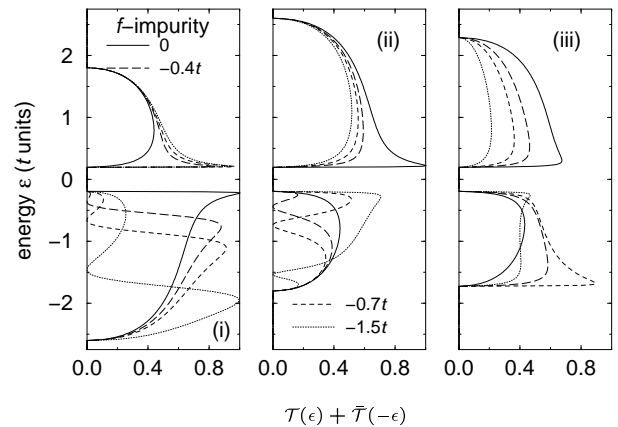,width=2.4in,,angle=90}}
\caption{
Total (electron plus hole) transmission coefficient
\protect{${\cal{T}}\!\left(\epsilon\right)
+\bar{\cal{T}}\!\left(-\epsilon\right)$}
in SCR as a function of the energy
$\epsilon$ (vertical axis). From left to right
vertical panels correspond to FE, NB, and SC,
respectively. Curves for different values of
the $f$-impurity energy $\varepsilon_{f1/2}$ are plotted.
A consequence
of equations (\ref{eq:canonical}) is that
\protect{${\cal{T}}\!\left(\epsilon\right)$} of FE for
 $\varepsilon_{f1/2}>0$
can be mapped into \protect{$\bar{\cal{T}}\!\left(-\epsilon\right)$} of NB
under the transformation \protect{$-\epsilon\rightarrow \epsilon$},
\protect{$-\varepsilon_{f1/2}\rightarrow \varepsilon_{f1/2}$}.
The parameters used are the same as in Fig.~\ref{fig2} with $T=0$.
}
\label{fig5}
\end{figure}
In order to gain some insight into the thermopower 
behavior, in Fig.~\ref{fig5} we have plotted the total transmission
coefficient \protect{${\cal{T}}\!\left(\epsilon\right)
+\bar{\cal{T}}\!\left(-\epsilon\right)$} vs $\epsilon$
for different values of 
$\varepsilon_{f1/2}$, similar to Fig.~\ref{fig4}.
FE and NB curves [panel (i) and (ii), respectively] are qualitatively
different from those of SC [panel (iii)],
as we set $\varepsilon_{f1/2}$ to 
negative values in the energy band range.
For FE and NB, $\bar{\cal{T}}$ goes to zero in the neighborhood of
$\varepsilon_{f1/2}$, as if the hole were completely backscattered from
the interface when resonates the energy with that of the impurity atom.
On the contrary, in case 
(iii) the effect is opposite, with $\bar{\cal{T}}$ gaining weight
for energies close to $\varepsilon_{f1/2}$. The trend is similar for
$\varepsilon_{f1/2}>0$, showing total reflection in the neighborhood of
positive values of $\varepsilon_{f1/2}$, in case (i) and (ii).
These results demonstrate that while the $f$-impurity
level of FE and NB does not share the electronic conduction, due to 
localization, it strongly affects the transmission, because either in case (i) 
the quasi-particle excitation is a coherent superposition of $d$- and 
$f$-states, or in case (ii) the ``bare'' bands are hybridized.
The case of SC is excluded here.

\begin{figure}
\begin{picture}(300,180)
\put(240,180){\epsfig{file=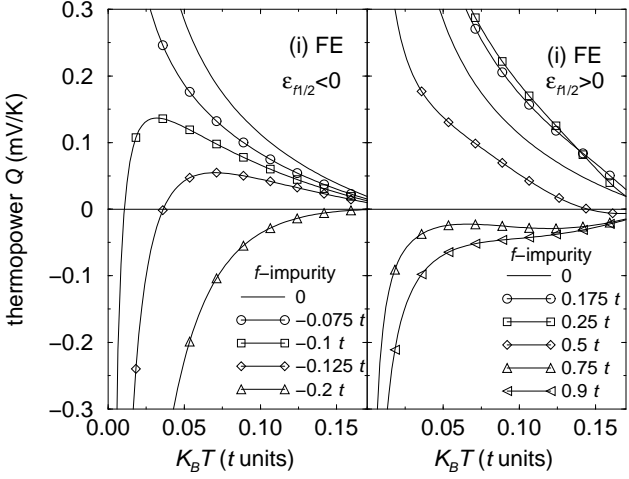,width=2.5in,,angle=-90}}
\end{picture}
\caption{
Thermopower $Q$ vs $T$ for FE. Left panel: Curves for different values
of the $f$-impurity energy $\varepsilon_{f1/2}$ (negative). 
Right panel: The same, with
\protect{$\varepsilon_{f1/2}>0$}. Note that $Q$ changes sign for some curves
as $T$ varies, and that \protect{$\left|Q\right|\rightarrow \infty$}
as \protect{$T\rightarrow 0$}. Bulk parameters as in Fig.~\ref{fig2}.
}
\label{fig6}
\end{figure}
The overall effect of the $f$-impurity layer on transport is so strong 
that it even changes the dominant (electron or hole) character of
the thermopower, i.e.~the sign of $Q$. For the sake of simplicity, we 
now consider only FE. In Fig.~\ref{fig6} we plot $Q$ vs $T$ for different 
values of $\varepsilon_{f1/2}$. Left panel refers to $\varepsilon_{f1/2}<0$.
We see that, as we set $\varepsilon_{f1/2}$ from zero to negative values,
$Q$ changes sign: already at $\varepsilon_{f1/2}=$ -0.1$t$
$Q$ has electron character ($Q<0$) as \protect{$T\rightarrow 0$},
while for $k_{\text{B}}T>0.02t$ $Q$ has hole character ($Q>0$). This 
behavior can be understood by examining 
Fig.~\ref{fig5}: the impurity level drastically
diminishes the weight of the transmission coefficient at the top of the 
valence band, while increasing it at the bottom of the conduction band, 
so that the sign of $Q$ is reversed
at low temperatures, where the only excited carriers are those
whose energies are close to the gap. Note that $Q$ is extremely
sensitive to the position of $\varepsilon_{f1/2}$, in contrast to the case
of the $d$-impurity layer.
The  right panel of Fig.~\ref{fig6} presents a similar situation for
$\varepsilon_{f1/2}>0$. Here the situation is a little less obvious: as
we raise the value of $\varepsilon_{f1/2}$ first $Q$ rises then drops
and changes sign. This is due to the asymmetry of electron and hole bands.
At first $\varepsilon_{f1/2}$, now positive, 
is close to the bottom of the conduction band,
favoring the hole transport because it depresses the 
thermally activated electronic channels.
Then, as $\varepsilon_{f1/2}$ is increased, weight is added to the 
transmission coefficient at the bottom
of the conduction band.
This weight is ``swept away'' from the energy neighborhood resonant
with $\varepsilon_{f1/2}$, that now is higher in energy with respect
to the band bottom.
This mechanism also makes $Q$ change sign.

\begin{figure}
\centerline{\epsfig{file=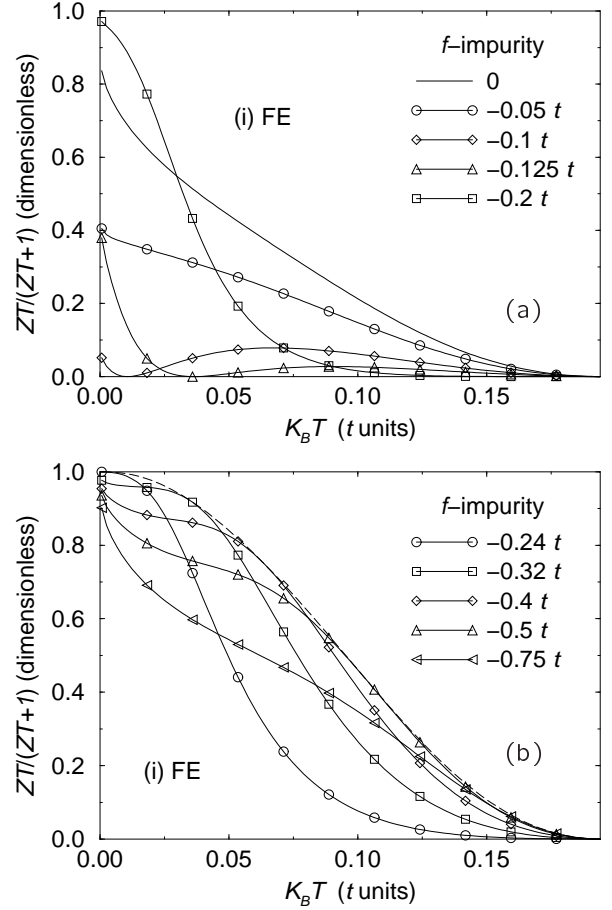,width=3.9in,,angle=0}}
\caption{
(a) \protect{$ZT/\left(ZT+1\right)$} vs temperature $T$ for different
values of the $f$-impurity level energy $\varepsilon_{f1/2}$ for FE.
Bulk parameters as in Fig.~\ref{fig2}.
(b) The same as Fig.~\ref{fig7}(a), for lower values of $\varepsilon_{f1/2}$.
Note that the ``shoulder'' in the curves is shifted towards higher values
of $T$, as $\varepsilon_{f1/2}$ is decreased.
The dashed line represents the envelope of the curves.
}
\label{fig7}
\end{figure}
To study the behavior of $ZT$ as the $f$-impurity  $\varepsilon_{f1/2}$
varies, we take a series of ``snapshots'' depicting 
\protect{$ZT/\left(ZT+1\right)$} vs $T$ for different values of
$\varepsilon_{f1/2}$ in Fig.~\ref{fig7}. Figure \ref{fig7}(a) focuses
on $\varepsilon_{f1/2}<0$, in a range
between 0 and -0.2$t$. We see that, as 
$\varepsilon_{f1/2}$ is lowered from its zero value,
\protect{$ZT/\left(ZT+1\right)$} conspicuously decreases: 
in this range $Q$ changes sign. At $\varepsilon_{f1/2}=$ -0.1 and -0.125$t$
a temperature exists at which $ZT=0$: these temperatures correspond to the
zero of $Q$ in Fig.~\ref{fig6} (left panel).
At $\varepsilon_{f1/2}=$ -0.2$t$, however, \protect{$ZT/\left(ZT+1\right)$}
begins to rise.
In Fig.~\ref{fig7}(b) we analyze the ``high-value region'' of 
\protect{$ZT/\left(ZT+1\right)$}. At $\varepsilon_{f1/2}=$ -0.24$t$ the 
curve is quite depressed around $k_{\text{B}}T=$ 0.1 
(room temperature with the 
usual numerical parameters), but it acquires giant values around $T=0$, 
close to the theoretical upper bound 1.
Moreover, the shape of the curve is very different from the typical pattern
of the $d$-impurity layer: 
in this latter case the curve is convex 
around $T=0$, while in the present situation 
\protect{$ZT/\left(ZT+1\right)$} is concave, 
i.e.~if one slightly departs from $T=0$ $ZT$ will still be very high.
This approximate flatness of the curve suggests that a stable 
low-temperature working point should exist for a junction-based device.
As $\varepsilon_{f1/2}$ is furtherly decreased, 
\protect{$ZT/\left(ZT+1\right)$}
shows a concave ``shoulder'' which is shifted towards higher temperatures.
This feature implies that, once we fix a temperature, 
it is possible to find
an optimal value of  $\varepsilon_{f1/2}$ maximizing the figure of merit.
This locus of optimal working points is manifestly given by the
envelope curve for the ``shoulders'' (dashed line).
The behavior of \protect{$ZT/\left(ZT+1\right)$} as $\varepsilon_{f1/2}$ rises
from zero ($\varepsilon_{f1/2}>0$) is similar, but 
the sequence is inverted with respect to the case we have just examined:
this time first 
\protect{$ZT/\left(ZT+1\right)$} dramatically increases, and then rapidly
drops to low values.

\begin{figure}
\begin{picture}(300,200)
\put(240,180){\epsfig{file=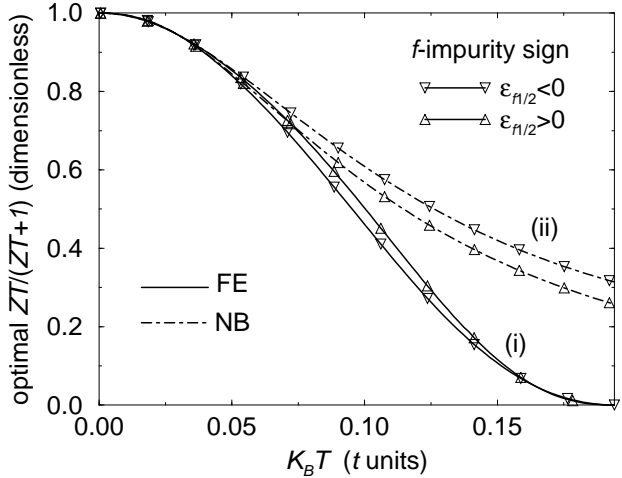,width=2.5in,,angle=-90}}
\end{picture}
\caption{Maximum value of
\protect{$ZT/\left(ZT+1\right)$} vs $T$ for an optimized choice
of the $f$-impurity level energy $\varepsilon_{f1/2}$.
Case (i) (solid lines) and (ii) (dot-dashed lines) are represented.
The up (down) triangles refer to the positive (negative) values of
local maxima $\varepsilon_{f1/2}$, computed at fixed $T$.
Note that the solid line with down triangles is the
envelope curve of Fig.~\ref{fig7}(b).
Bulk parameters as in Fig.~\ref{fig2}.
}
\label{fig8}
\end{figure}
Figure \ref{fig8} shows 
the maximum values of \protect{$ZT/\left(ZT+1\right)$}
vs $T$ that can be reached with a suitable $f$-impurity energy
$\varepsilon_{f1/2}$. Up (down) triangles indicate positive (negative)
optimal energies $\varepsilon_{f1/2}$.
Analogous results for NB (dot-dashed lines) are shown.
The curves correspond to the loci of the $T$-$\varepsilon_{f1/2}$ space 
like the envelope curve of Fig.~\ref{fig7}(b).
From Fig.~\ref{fig8} we can derive the absolute maxima of $ZT$. 
Our results are that, at room temperature ($k_{\text{B}}T=$ 0.1$t$, with the
usual numerical parameters) \protect{$ZT\sim$ 1} for FE and 1.6 for NB,
at \protect{$T=$ 150 K} $ZT$ is 
\protect{$\sim$ 6} in both cases, at \protect{$T=$ 100 K}
is \protect{$\sim$ 13}, 
and at \protect{$T=$ 40 K} is \protect{$\sim$ 100}.

These extremly high values suggest the possibility of engineering 
a periodic lattice of $\delta$-layers made of metal
and strongly correlated semiconductor to create an efficient 
thermoelectric device. 
Borrowing an intuition of Refs.~\onlinecite{thermionic} and 
\onlinecite{moyzhes},
the bias applied to the superlattice should be such that
the electronic transport occurs perpendicular to interfaces.
The thickness of each layer should be comparable to the electronic mean
free path, and large enough to prohibit electrons from tunnelling.
In this regime the electronic motion within
each layer is ballistic, and the interfaces can be considered as the
only scattering mechanisms.
Mahan and Woods\cite{thermionic} call thermionic a similar device
with an ordinary semiconductor replacing the strongly
correlated one, because
the electronic motion is ballistic and the
expression for the current is the Richardson's equation. However, here the
situation is different, because the transmission coefficient
has such a sharp variation with energy
that the current has a more complex form than the Richardson's expression
and it requires a full analysis of the role of the interface in transport.
The idea here is to grow a $f$-atom overlayer 
at $z=a/2$ with optimal $f$-level energy $\varepsilon_{f1/2}$ 
depending on the working temperature of the device.
In addition, we find that
the magnitude of thermal conductance changes only slightly
with $\varepsilon_{f1/2}$.
The $\delta$-layer of rare-earth atoms could also act
like an additional source of strong scattering for phonons, 
thus lowering the lattice
thermal conductance.\cite{physicstoday} Therefore the junction is
a promising candidate as a thermoelectric device. 
 
\subsection{Effects of the surface on the lattice 
structure}\label{rilassamento}

We briefly comment on the influence which the interface has on the 
ideal lattice structure. We consider the narrow band semiconductor NB. 
If the effect of the surface at $z=0$ is to locally shorten 
the lattice constant $a$, this will certainly change the hybridization 
parameter $\tilde{V}_{dfL}$ between the $d$-site at $z=0$ and the $f$-site 
at $z=a/2$. Presumably $\tilde{V}_{dfL}$ will be increased. 
Modifying nothing but $\tilde{V}_{dfL}$ with respect to the clean 
interface case, we find that $ZT$ is only slightly affected by this
relaxation effect. In particular, to change $\tilde{V}_{dfL}$ up to
10\% almost rigidly shifts $ZT/(ZT+1)$ over the whole temperature
range by approximately 10 -- 15 \%.
From a check of the 
transmission coefficient, we find
that this surface effect represents only a minor perturbation to the
electronic transport across the junction.

\section{Conclusions}\label{fine}

We have made a qualitative theoretical study of the possibility that a
junction of metal and FE or NB (as opposed to bulk materials or to
a junction metal/SC) produces high thermopower.
This is possible if a $\delta$-layer of
suitable rare-earth impurity atoms substitutes
the original layer at the interface. The localized 
character of the impurity $f$-orbital has a strong effect
on the transmission of carriers across the junction.
The figure of merit $ZT$ attained is very high, especially at low 
temperatures. A realistic device is proposed which exploits the
thermoelectric potentialities of the junction.

\begin{acknowledgments}
This work is supported by the NSF contract DMR 0099572,
MIUR-FIRB ``Quantum phases of ultra-low electron density semiconductor
heterostructures'', and MIUR Progetto Giovani Ricercatori. 
The authors thank Professors
D.~Chemla, M.~L.~Cohen and S.~G.~Louie for the hospitality 
of University of California Berkeley, where this work was initiated.
M.~R.~is grateful to Professors Franca Manghi and Elisa Molinari for all the 
help given. L.~J.~S.~thanks J.~E.~Hirsch for helpful discussions
and the Miller Institute for support.
Con il contributo del Ministero degli Affari Esteri,
Direzione Generale per la Promozione e la Cooperazione
Culturale.
\end{acknowledgments}

\appendix

\section{Current operator}\label{appcurrent}

In this appendix 
we derive the form of the density current operator $J$ for the effective
Hamiltonian ${\cal{H}}^{\rm MF}$ of  
Eq.~(\ref{eq:h})
(or ${\cal{H}}^{\rm MF}-\mu$). This is a key quantity in computing the
transport properties of the junction.

One way to find $J$ is to exploit the gauge invariance of the hamiltonian 
(\ref{eq:h}). In the presence of an oscillating, uniform electric 
field $E\!\left(t\right)=E{\rm e}^{-{\rm i}\omega t}$ 
along $z$ ($t$ is the time
and $\omega/2\pi$ is the frequency), one must add to
${\cal{H}}^{\rm MF}$ a term representing the coupling to the applied field.
One thus introduces the vector potential \protect{$A\!\left(t\right)=
cE{\rm e}^{-{\rm i}\omega t}/{\rm i}\omega$} such that 
\protect{$E\!\left(t\right)=-\dot{A}/c$}, where $c$ is the speed of light.
The gauge transformation
\begin{eqnarray}
d_j^{\prime} &=& d_j\exp{\left[-\frac{{\rm i}e}{\hbar c}A\!\left(t\right)aj
\right]},
\nonumber\\
f_{j+1/2}^{\prime} &=& f_{j+1/2}\exp{\left[-\frac{{\rm i}e}{\hbar c}A\!
\left(t\right)a\left(j+1/2\right)\right]},
\label{eq:gauge}
\end{eqnarray}
removes the coupling to the vector potential $A\!\left(t\right)$ 
preserving the gauge invariance of ${\cal{H}}^{\rm MF}$;
if we apply the transformation (\ref{eq:gauge}) to ${\cal{H}}^{\rm MF}$
we obtain a new Hamiltonian  ${{\cal{H}}^{\rm MF}}^{\prime}$ which must be
given, expanding to first order in $A$, by
\begin{equation}
{{\cal{H}}^{\rm MF}}^{\prime}\sim {\cal{H}}^{\rm MF} -\frac{aJ}{c}
A.
\label{eq:defJ}
\end{equation}
Equation (\ref{eq:defJ}) permits to identify $J$, whose expression is
\begin{equation}
J=\sum_j J\!\left(j\right),
\end{equation}
\begin{equation}
 J\!\left(j\right)=\frac{\rm i}{\hbar}et^{\prime}\,
d_{j+1}^{\dagger}d_j {\rm \;+\; H.c.}\quad
\forall j<0,
\label{eq:jless}
\end{equation}
\begin{eqnarray}
 J\!\left(0\right) &=& \frac{\rm i}{\hbar}et\,
d_1^{\dagger}d_0+\frac{{\rm i} e}{2\hbar}\Big[\left(\tilde{\Delta}_L
+\tilde{V}_{dfL}\right)f_{1/2}^{\dagger}d_0 \nonumber\\
&&-\quad\left(\tilde{\Delta}_R
-\tilde{V}_{dfR}\right)f_{1/2}^{\dagger}d_1\Big]{\rm \;+\; H.c.},
\end{eqnarray}
\begin{eqnarray}
&&  J\!\left(j\right)=\frac{\rm i}{\hbar}et\,
d_{j+1}^{\dagger}d_j+\frac{{\rm i} e}{2\hbar}\Big[\left(\Delta
+V_{df}\right)f_{j+1/2}^{\dagger}d_j
\nonumber\\ && -\quad\left(\Delta-V_{df}\right)
f_{j+1/2}^{\dagger}d_{j+1}\Big]{\rm \;+\; H.c.}\quad\forall j>0.
\end{eqnarray}
Here $J\!\left(j\right)$ is the electric current operator at atomic site $j$
(in one dimension the current coincides with its density).
If the transport across the junction is stationary, at each site $j$ 
the current must be conserved; thus,
to identify the density current associated 
with each quasi-particle excitation,
we can calculate the operator $J\!\left(j\right)$ at some
suitable atomic site $j$ where its expression is simpler:
this is used in Sec.~\ref{formalism}.
For $j<0$ ${\cal{H}}^{\rm MF}$ is a simple tight-binding Hamiltonian,
and the formula for $J\!\left(j\right)$
becomes, replacing $d$ and $f$ operators in (\ref{eq:jless}) with
$\gamma$ operators:
\begin{eqnarray}
&&J\!\left(j\right)=\frac{2et^{\prime}}{\hbar N_s}\sum_k {\rm Im}\left[
u_k^*\!\left(j\right)u_k\!\left(j+1\right)\right]\gamma^{\dagger}_{ke}
\gamma_{ke}
\nonumber\\
&&-\quad\frac{2et^{\prime}}{\hbar N_s}\sum_k{\rm Im}\left[
\bar{u}_{-k}\!\left(j\right)\bar{u}^*_{-k}\!\left(j+1\right)\right]
\gamma^{\dagger}_{-kh}\gamma_{-kh}\nonumber\\
&&\quad+\qquad\frac{2et^{\prime}}{\hbar N_s}
\sum_k{\rm Im}\left[\bar{u}_{-k}
\!\left(j\right)\bar{u}^*_{-k}\!\left(j+1\right)\right].
\label{eq:jtb}
\end{eqnarray}
It turns out that the constant third term on the right-hand side of 
Eq.~(\ref{eq:jtb}) is always zero, 
hence it is easy to identify the electric current 
$J_{ke}\!\left(j\right)$ and $J_{-kh}\!\left(j\right)$ 
associated with each electron or hole excitation, respectively:
\begin{eqnarray}
J\!\left(j\right)&=&\frac{1}{N_s}\sum_k\Big[J_{ke}\!\left(j\right)
\gamma^{\dagger}_{ke}\gamma_{ke}
\nonumber\\ &&+\quad J_{-kh}\!\left(j\right)
\gamma^{\dagger}_{-kh}\gamma_{-kh}\Big],
\label{eq:counting}
\end{eqnarray} 
\begin{equation}
J_{ke}\!\left(j\right)=\frac{2et^{\prime}}{\hbar}
{\rm Im}\left[u_k^*\!\left(j\right)u_k\!\left(j+1\right)\right]
\quad \forall j<0,
\label{eq:je}
\end{equation}
\begin{eqnarray}
J_{-kh}\!\left(j\right) &=& -\frac{2et^{\prime}}{\hbar}
{\rm Im}\Big[\bar{u}_{-k}\!\left(j\right)
\nonumber\\ && \times\quad \bar{u}^*_{-k}
\!\left(j+1\right)\Big]\quad \forall j<0.
\label{eq:jh}
\end{eqnarray}
The particle current density for electrons 
\protect{$J_{keN}\!\left(j\right)$} is given by
\begin{equation}
J_{keN}\!\left(j\right)=\frac{J_{ke}\!\left(j\right)}{e},
\label{eq:enumber}
\end{equation}
and for holes \protect{$J_{-khN}\!\left(j\right)$} by 
\begin{equation}
J_{-khN}\!\left(j\right)=\frac{J_{-kh}\!\left(j\right)}{-e}.
\label{eq:hnumber}
\end{equation}

The picture consistent with these results is that 
$\left(u_k\!\left(j\right),v_k\!\left(j+1/2\right)\right)$ [or 
$\left(\bar{u},\bar{v}\right)$] represents the two-component electron
(hole) wavefunction of the elementary excitation, and
$J_{keN}\!\left(j\right)$ ($J_{-khN}$) is its probability
current density. Indeed, equations (\ref{eq:je}-\ref{eq:jh}) can be
derived also from the continuity equation of wavefunctions,
if one
associates the electron amplitude $\left(u,v\right)$ with the charge
$e$, and the hole amplitude $\left(\bar{u},\bar{v}\right)$
with $-e$. To obtain the continuity equation, one has to multiply
the time-dependent BdG equation (\ref{eq:BdGbisstar}) [(\ref{eq:BdGbisend})]
by $u^*_k(j,t)$ [$\bar{u}^*$],
and the complex conjugate of (\ref{eq:BdGbisstar}) [(\ref{eq:BdGbisend})]
by $u_k(j,t)$ [$\bar{u}$], and finally subtract one term from the other.

In an analogous way we can also write the expression of the
current for $j>0$. If $\Delta$ and $V_{df}$ are real, one obtains:
\begin{eqnarray}
&& J_{ke}\!\left(j\right)=\frac{2et}{\hbar}
{\rm Im}\left[u_k^*\!\left(j\right)u_k\!\left(j+1\right)\right]
 + \frac{e}{\hbar}\left(\Delta+V_{df}\right)
\nonumber\\
&&\quad\times\quad{\rm Im}\left[
u_k^*\!\left(j\right)v_k\!\left(j+1/2\right)\right]
-\frac{e}{\hbar}\left(\Delta-V_{df}\right) \nonumber\\
&&\qquad \times\quad {\rm Im}\left[
u_k^*\!\left(j+1\right)v_k\!\left(j+1/2\right)\right],
\label{eq:ecurrentbulk}
\end{eqnarray}
\begin{eqnarray}
&& J_{-kh}\!\left(j\right)=-\frac{2et}{\hbar}
{\rm Im}\left[\bar{u}_{-k}\!\left(j\right)\bar{u}^*_{-k}
\!\left(j+1\right)\right]-\frac{e}{\hbar}\left(\Delta+V_{df}\right) \nonumber\\
&& \quad\times\quad {\rm Im}\left[\bar{u}_{-k}\!\left(j\right)\bar{v}^*_{-k}\!\left(
j+1/2\right)\right]
+\frac{e}{\hbar}\left(\Delta-V_{df}\right) \nonumber\\
&& \qquad\times\quad
{\rm Im}\left[\bar{u}_{-k}\!\left(j+1\right)\bar{v}^*_{-k}\!\left(
j+1/2\right)\right].
\label{eq:hcurrentbulk}
\end{eqnarray}


\clearpage
\clearpage
%
%
%


\begin{thebibliography}{}
\bibitem{gaugeinvariance}
When $V_{df}$ is exactly zero, the local gauge invariance of the
Falicov-Kimball Hamiltonian renders the $f$ electron site occupation
number classical. See J.~K.~Freericks and V.~Zlati\'c\cite{Freericks}
and references therein. However, a small $V_{df}$ breaks the
local gauge invariance and the long-range order of FE is 
possible.\cite{ferro} The broken local gauge invariance
by hopping $f$-electrons is shown beyond mean-field approximation
to lead to a rich phase diagram including the FE phase.\cite{Batista}
\bibitem{kappa}
Here $\kappa$ should be the total thermal conductivity: however,
we compute only the electronic contribution.
\end{thebibliography}
\end{document}